\documentstyle[emulateapj]{article}
\tighten
\newcommand{\etal}{{\rm et al.~}}
\newcommand{\Mpc}{$h^{-1}$~{\rm Mpc}}
\newcommand{\hmpc}{$h$~{\rm Mpc$^{-1}$}}

\def\apj{ApJ\ }
\def\apjl{ApJL\ }  
\def\apjs{ApJS\ }
   
\def\mn{MNRAS }

\begin{document}

\title{Optical and X-ray clusters  as tracers of the supercluster-void
network. III Distribution of Abell and APM clusters }

\author{ Maret Einasto\altaffilmark{1}, 
Jaan Einasto\altaffilmark{1}, 
Erik Tago\altaffilmark{1}, 
Heinz Andernach\altaffilmark{2}, 
Gavin B.  Dalton\altaffilmark{3},
and Volker M\"uller\altaffilmark{4} }

\altaffiltext{1} {Tartu Observatory, EE-61602 T\~oravere, Estonia}
\altaffiltext{2} {Depto.  de Astronom\'\i a, Univ.\ Guanajuato,
            Apdo.\ Postal 144, Guanajuato, C.P.\ 36000, GTO, Mexico}
\altaffiltext{3} {University of Oxford, Department of Astrophysics, 
Keble Road, Oxford, OX1 3RH, U.K.}
\altaffiltext{4} {Astrophysical Institute Potsdam, An der Sternwarte 16,
       D-14482 Potsdam, Germany}

\begin{abstract} 

We present a comparison of Abell and Automated Plate Measuring
Facility (APM) clusters of galaxies as tracers of the large scale
structure of the Universe.  We investigate selection functions of both
cluster catalogs, using samples of all clusters (including clusters
with estimated redshifts), and samples of clusters with measured
redshifts.  We find that the distribution of rich superclusters,
defined by all Abell and APM clusters, is similar in volume covered by
both cluster samples. We show that the supercluster-void network can
be traced with both cluster samples; the network has a well-defined
period of $\sim 120$~\Mpc. We calculate the correlation function for
Abell and APM cluster samples.  However, the APM cluster sample with
measured redshifts covers a small volume which contains only a few
very rich superclusters.  These superclusters surround one void and
have exceptionally large mutual separations. Due to this property the
secondary maximum of the correlation function of APM clusters with
measured velocities is located at larger scales than corresponding
feature in the correlation function of Abell clusters.  We conclude
that the APM sample is not representative for the large-scale
structure as a whole due to the small space coverage.  The Abell
cluster catalog is presently the best sample to investigate the
large-scale distribution of high-density regions in the Universe.  We
present a catalog of superclusters of galaxies, based on APM clusters
up to a redshift $z_{lim}=0.13$.

\end{abstract}


\section {Introduction}

In the present series of papers we study the properties of the
supercluster-void network as delineated by different types of clusters
of galaxies, as well as the regularity of this network.  So far the
supercluster-void network has been studied in detail only using
samples of Abell clusters of galaxies (Abell 1958, Abell, Corwin,
Olowin 1989, hereafter ACO).  In the first two papers of this series
we compared the distribution of Abell clusters with clusters selected
on the basis of their X-ray emission (Einasto \etal 2001, Paper I) and
derived their correlation functions (Tago \etal 2001, Paper II).  Here
we continue the comparison of samples of clusters of galaxies as
tracers of the supercluster-void network, and now use the cluster
catalog derived from scans with the Automated Plate Measuring (APM)
Facility (Dalton \etal 1997, hereafter D97) based on the APM galaxy
catalog (Maddox, Efstathiou \& Sutherland 1996).  One reason to
compare the Abell and APM samples of clusters is the fact that power
spectra derived from these cluster samples are different from each
other.  The power spectrum of Abell clusters has a sharp peak around
the scale $l \approx 120$~\Mpc\ or wavenumber $k=2\pi/l = 0.05$~\hmpc\
(Einasto \etal 1997a, hereafter E97a, Retzlaff \etal 1998, hereafter
R98).  In contrast, the maximum of the power spectrum of APM clusters
occurs on larger scales, $k \approx 0.04$~\hmpc, and it is flatter
(Tadros \etal 1998, hereafter T98).  The reason for these
discrepancies is not fully clear.  An independent study of Miller \&
Batuski (2000) has shown only a mild feature on a scale $k =
0.05$~\hmpc, and the amplitude of the power spectrum increases toward
larger scales.

The main goal of this paper is to compare the distribution of Abell
and APM clusters of galaxies and to find how well these cluster
catalogs can be used to investigate properties of the
supercluster-void network.  We shall use a cosmographic approach to
describe superclusters and voids; we shall also apply various methods
to quantify the distribution to allow the comparison of observations
with theory.  

The paper is organized as follows.  In Section~2 we describe
the cluster catalogs used: the latest compilation of redshifts of Abell
clusters by Andernach \& Tago (1998), and the published catalog of APM
clusters by D97.  We derive selection functions for both samples.  In
Section~3 we find the multiplicity function of superclusters in the
APM catalogs, compile a supercluster catalog for the APM sample, and
compare the distribution of rich superclusters in the Abell and APM
catalogs; a supercluster catalog based on Abell sample was compiled in
Paper I.  To characterize the distribution of clusters in quantitative
terms in Section~4 we derive the correlation function and power
spectrum; we also investigate the distribution of neighbors of
superclusters.  We discuss our results and compare them with results
of other studies in Section~5.  Section~6 summarizes the main
conclusions.  In the Appendix we give a catalog of superclusters
constructed on the basis of APM clusters, using a redshift limit
$z_{lim}=0.13$; it is also available electronically at the web site
http://www.aai.ee.

\begin{figure*}[ht] 
\vspace*{10cm} 
\figcaption{The distribution of Abell and APM clusters
on the sky in equatorial Aitoff coordinates, plotted as open circles
and crosses, respectively.  Declination $\delta$ and right ascension
$\rlap{b}$ are given in degrees, both for the equinox 1950.  To avoid
splitting of the region covered by APM sample, the zero point of the
right ascension is put in the middle of the plot.  }
\includegraphics{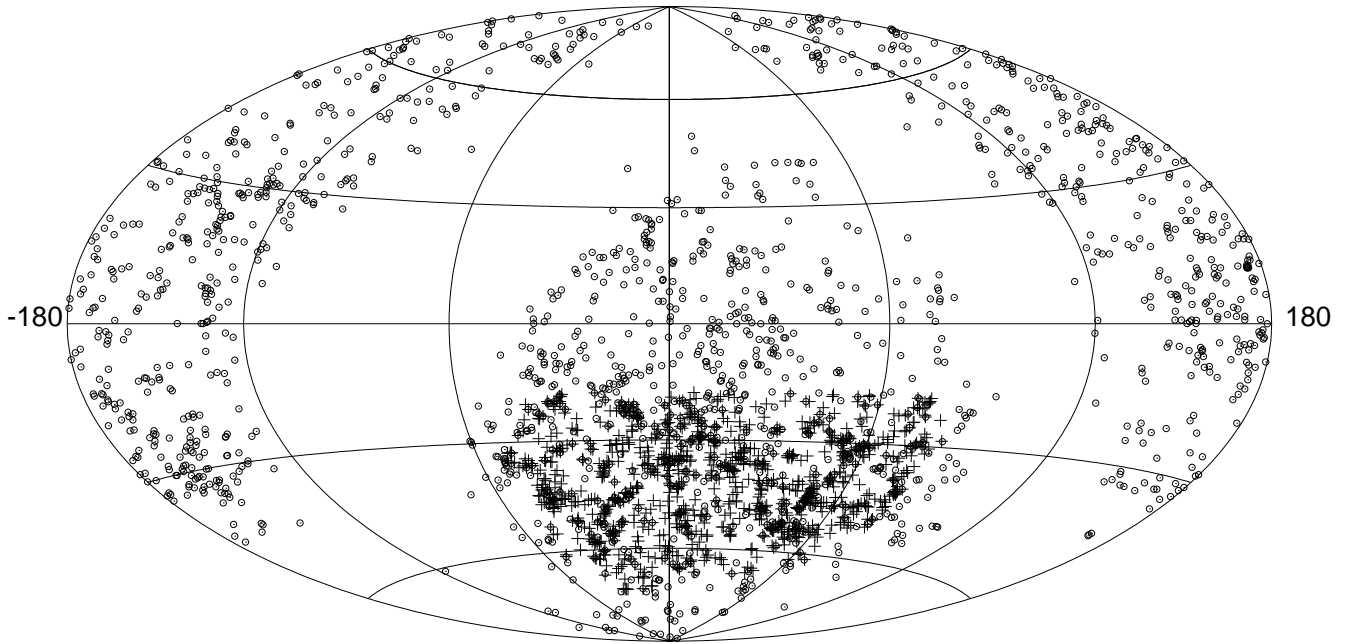}
\label{figure1}
\end{figure*}

\section{The comparison of Abell and APM cluster catalogs}

\subsection{Selection effects in Abell and APM cluster catalogs}

For the present study we use the version of March 1999 of the redshift
compilation for Abell clusters described by Andernach \& Tago (1998).
This compilation contains all known measured redshifts of Abell
clusters, based on redshifts of individual cluster galaxies, and
redshift estimates on the basis of the 10th brightest member of the
cluster according to the formula derived by Peacock \& West(1992), for
both Abell catalogs (Abell 1958 and ACO).  We omitted from the
compilation all supplementary, or S-clusters, but included clusters of
richness class 0.  From this general sample we selected all clusters
with measured redshifts not exceeding $z_{lim}=0.15$.  If no measured
redshift was available we applied the same criterion for estimated
redshifts. This sample contains 2276 clusters, 1258 of which have
measured redshifts.  We consider that a cluster has a measured
redshift if at least one of its member galaxy has a measured redshift.
In cases where the cluster has less than three galaxies with measured
redshifts, and the measured and estimated redshifts differ by more
than a factor of two ($|\log (z_{meas}/z_{est})| > 0.3$,) the
estimated redshift was used.  In the case of component clusters (A,B,C
etc) with comparable number of measured redshifts, we used only the
cluster which better matches the estimated redshift.

The APM cluster catalog by D97 contains 957 clusters, 374 of which
have measured redshifts.  This cluster catalog was selected using the
apparent magnitude limit, $m_X=19.4$, as a distance indicator.  This
distance indicator is similar to Abell's magnitude $m_{10}$ of the
10th brightest galaxy in the cluster, used as distance indicator for
the Abell catalog.  In the case of the APM catalog the rank $X=C_i$ is
calculated from the sum of weights of galaxies brighter than $m_i$;
$i$ is the rank of the galaxy in the list ordered by magnitude;
$C_i=X=10$ is equivalent to Abell's rank $m_{10}$ (for details see
D97).  The apparent magnitude limit $m_X=19.4$ corresponds to a limit
in the estimated redshift, $z_{est}=0.118$.

\begin{figure*}[ht] 
\vspace*{7cm}
\figcaption {The relation between the measured and estimated redshift
for Abell and APM cluster samples. The straight line is $z_{meas}= z_{est}$.
}
\includegraphics{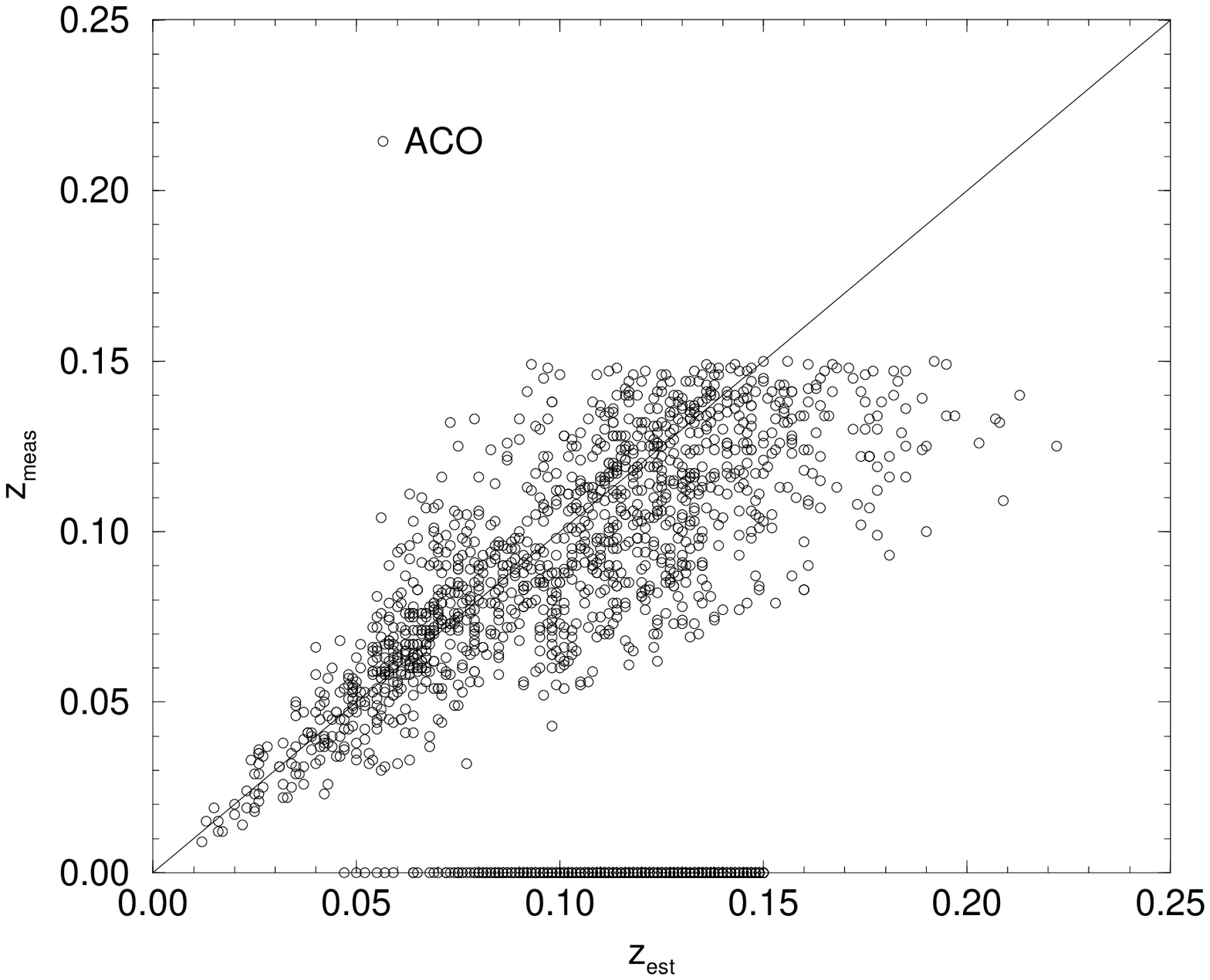}
\includegraphics{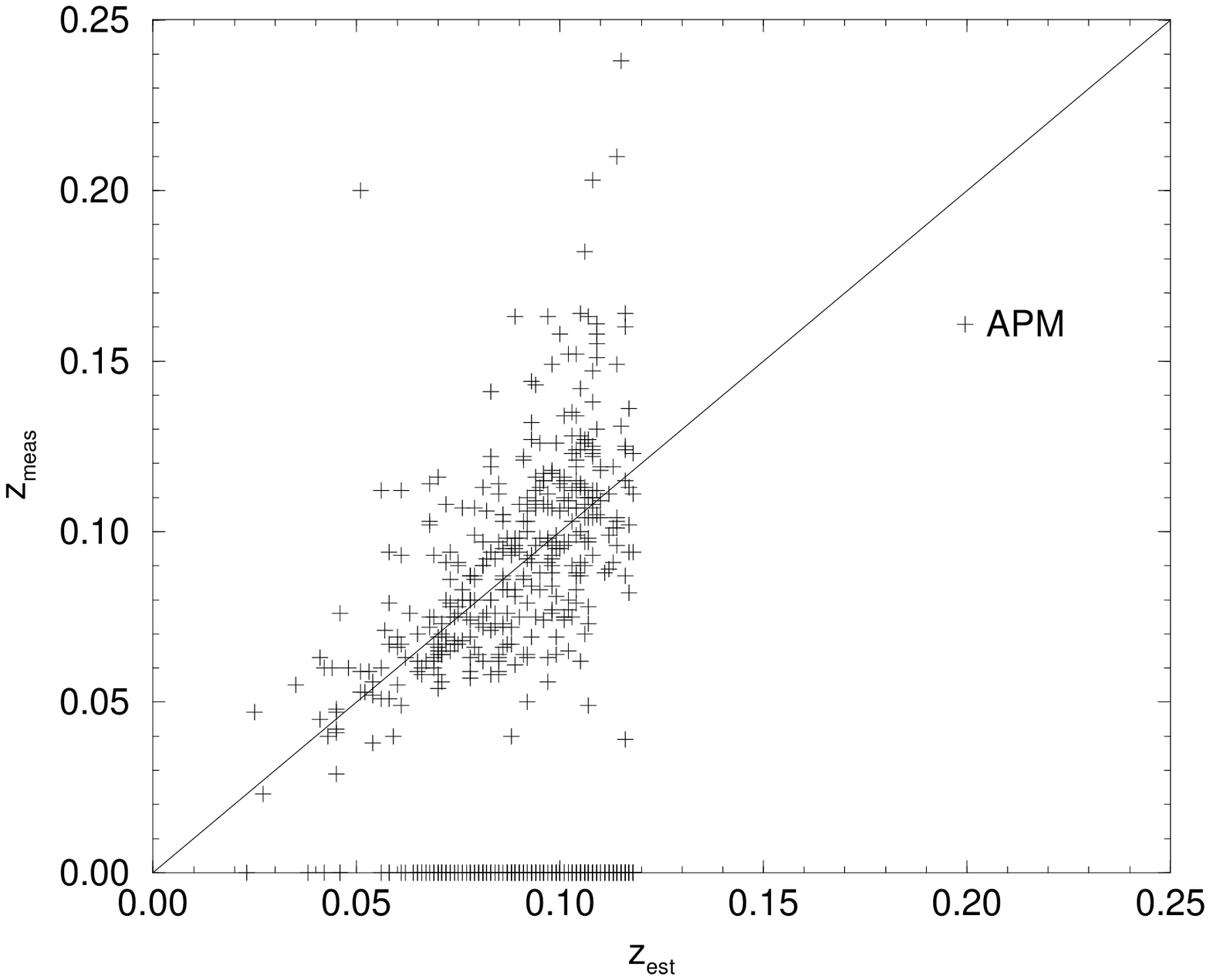}
\label{figure2}
\end{figure*}

The distribution of both ACO and APM clusters of galaxies on the sky
is shown in Figure~1.  We see that the APM cluster sample covers an
area much smaller than the area covered by Abell clusters.  We shall
estimate the effective area and volume of both cluster samples in
Section 5.2.  The decrease of the surface density of Abell clusters
towards the Galactic equator is well seen, and later in this Section
we determine this dependence quantitatively.  For the APM sample the
surface density of clusters seems to be constant. However, as our
analysis shows, here the surface density also depends on the Galactic
latitude.
 
As a first step to investigate the selection effects in both cluster
catalogs we looked for the relationship between estimated and measured
redshifts (Figure~2). As noted above, the Abell cluster sample was
selected using an upper limit of measured redshifts, $z_{lim}=0.15$;
if, instead, no measured redshift were available, a similar upper
limit was applied for estimated redshifts.  An inspection of Figure~2
shows that there exists a population of clusters with $z_{meas} <
0.15$ and $z_{est} > 0.15$.  This is due to errors in estimated
redshift. Thus it is natural to expect that some clusters within our
distance limit have estimated redshifts above this limit.  The
preponderance of points {\em below} the line $z_{meas}= z_{est}$ is
due to a preferred assignation of foreground galaxies to the cluster
as compared to background galaxies. 

The APM cluster sample was constructed in a different manner.  Here
clusters were preselected for the redshift observing program using an
{\em upper limit for the estimated redshift}, $z_{lim}=0.118$.  Since
estimated redshifts are subject to random errors, some clusters of
this sample have measured redshifts exceeding this limit.  Similarly,
some clusters that are located within the sphere of radius $z=0.118$,
are not included in the observing program since their estimated
redshifts are larger than their true redshifts.  This causes a large
deficit of clusters near the far side of the sample, as we see below.

\begin{figure*}[ht]
\vspace*{7cm}
\figcaption{Selection function in distance; left-hand panel: Abell
clusters; right-hand panel: APM clusters.  Bold lines show the
distribution of all clusters, thin lines the distribution of clusters
with measured redshifts.  Solid lines are for clusters in
superclusters of all richness classes (including isolated clusters),
dashed and dot-dashed lines are for clusters, located in superclusters
of multiplicity $N_{cl} \ge 4$ and $N_{cl} \ge 8$, respectively.  
}
\includegraphics{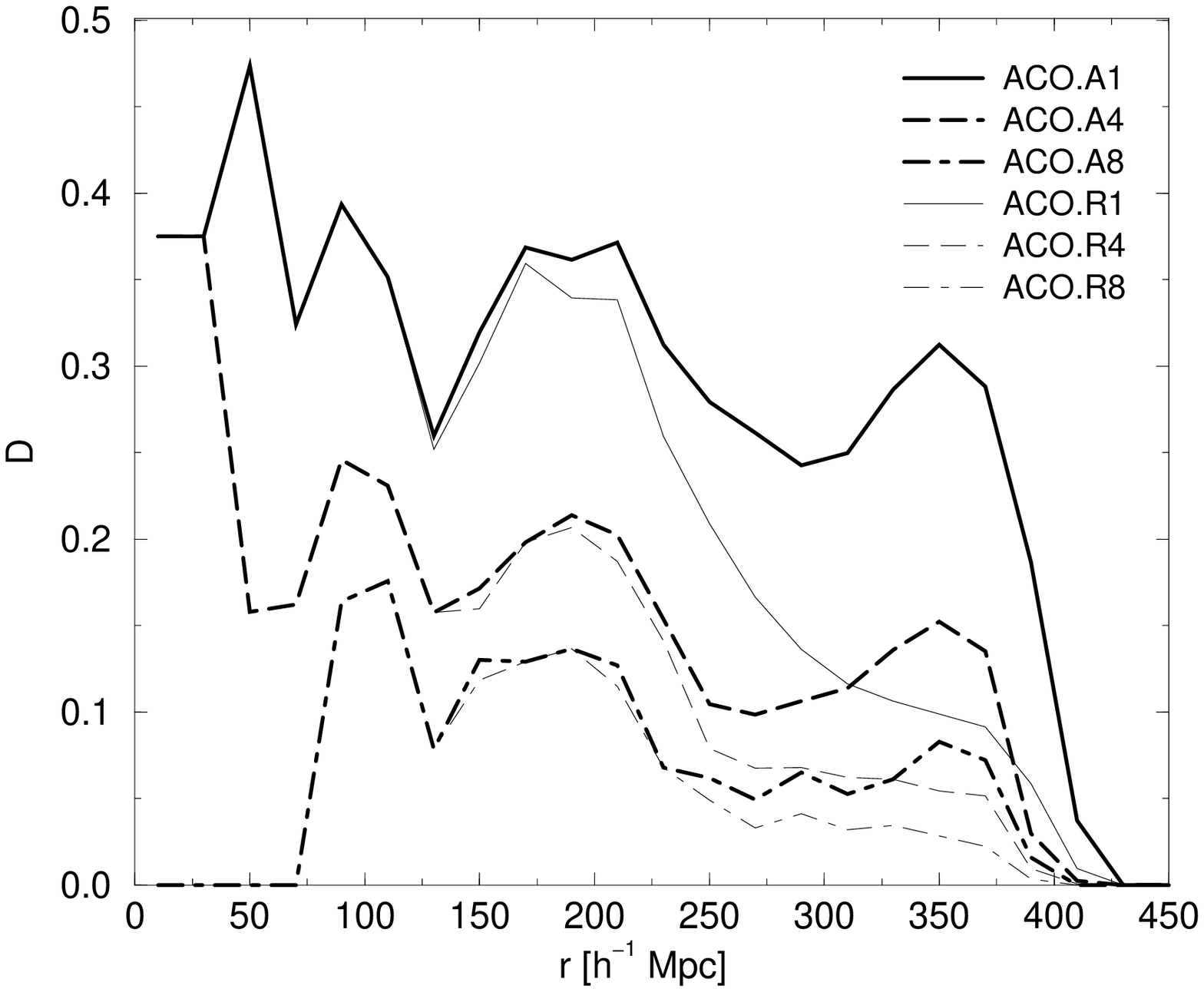} 
\includegraphics{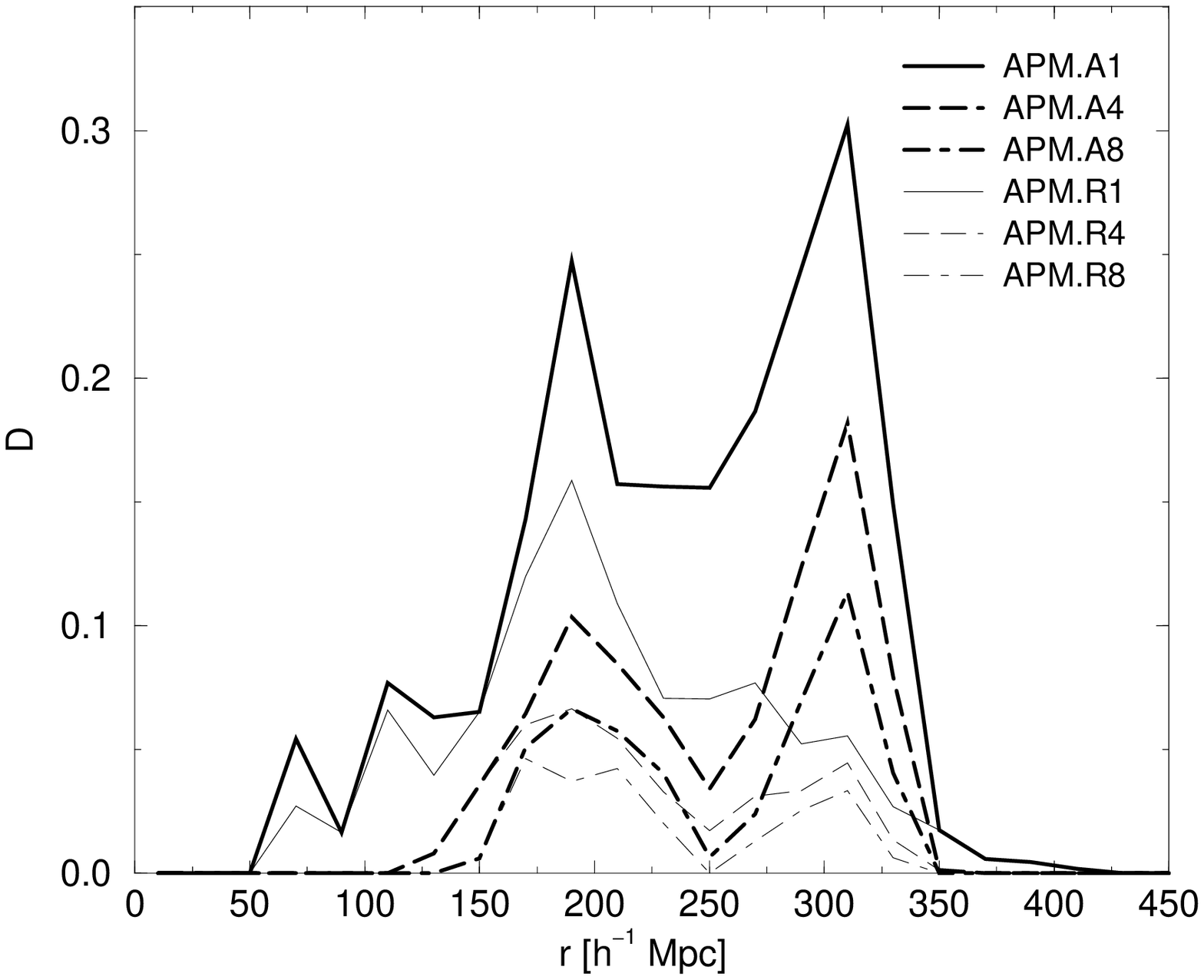}
\label{figure3}
\end{figure*} 

\begin{figure*}[ht]
\vspace*{7cm} 
\figcaption{Selection function in Galactic latitude.  Left hand panel:
Abell clusters; right hand panel: APM clusters.  Designations are
as in  Figure~3. The bold short-dashed line in the right panel shows the
selection function for a model that covers the region observed by APM
cluster survey with homogeneous surface density of test points.  }
\includegraphics{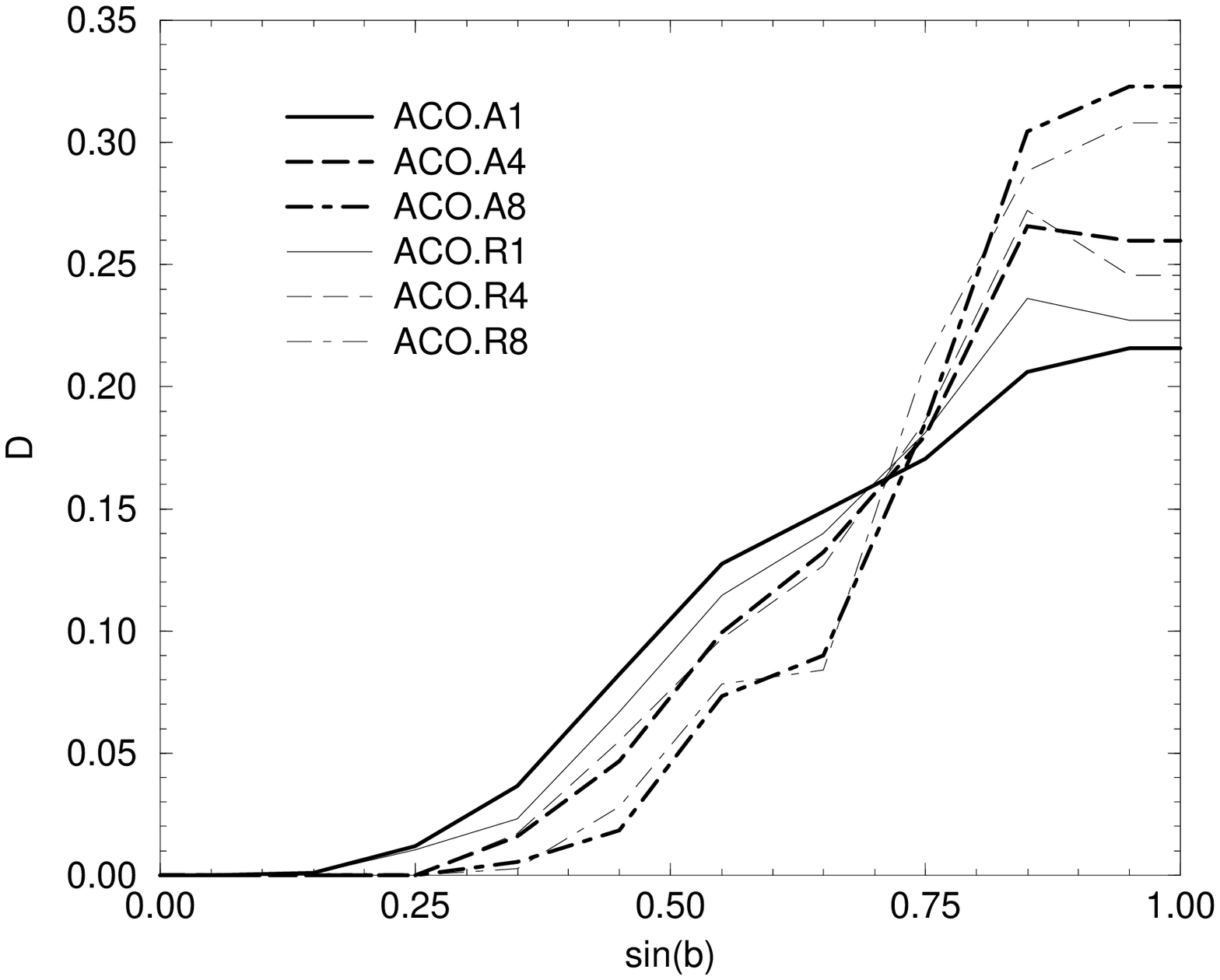} 
\includegraphics{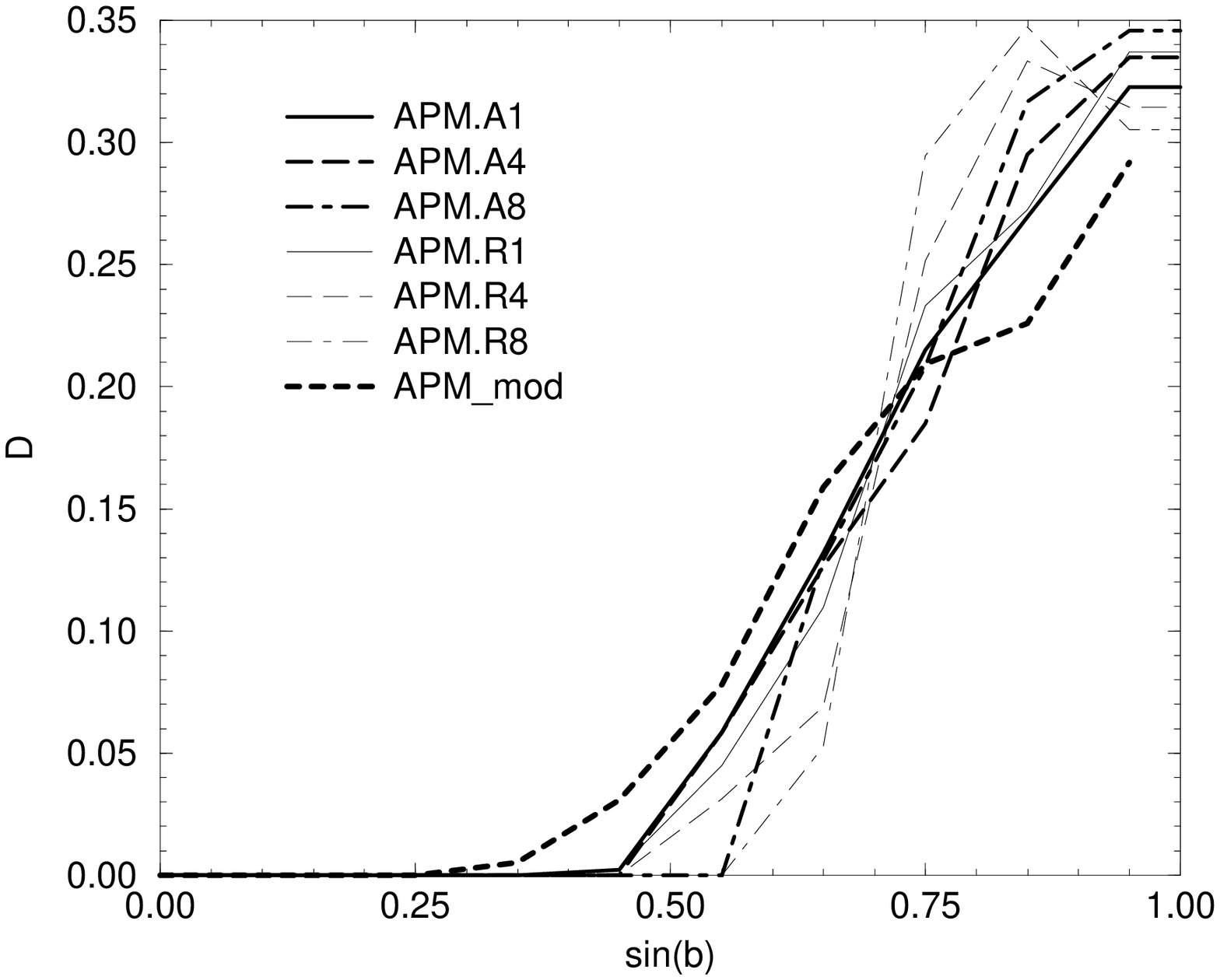}
\label{figure4}
\end{figure*}

We convert the redshifts into radial distances $r$ using the formula
by Mattig (1958) as in Paper I.  We adopt $H_0 =
h~100$~km~s$^{-1}$~Mpc$^{-1}$, and $q_0=\Omega_0/2=0.5$.  $\Omega_0$
is the density parameter (baryonic and dark matter) in units of the
critical density.

In Figure~3 we plot the spatial density of clusters of galaxies in
concentric shells of thickness 20~\Mpc\ in arbitrary units. Densities
were calculated from the number of clusters in respective shells,
divided by the volume of the shell (including regions not observed,
thus this diagram cannot be directly used to calculate absolute
spatial densities of clusters).  To avoid a very small test volume in
our vicinity, the first shell is taken between radii $r=0$ and
$r=40$~\Mpc.  We are interested in the spatial distribution of
clusters in superclusters of different richness, which is defined as
the number of clusters in superclusters, $N_{cl}$.  Thus we plotted
densities of clusters that belong to superclusters using a lower limit
of supercluster richness, $N_{cl} \ge 1$ (all clusters of the sample,
including isolated clusters), $N_{cl} \ge 4$, and $N_{cl} \ge 8$,
i.e. clusters in superclusters with at least 4 or 8 members (E97d).
Supercluster catalogs shall be discussed in the next subsection.
They were constructed separately for all clusters, and for clusters
having measured redshifts, both for the Abell and APM cluster samples.

Figure~3 shows that the spatial density of all clusters in the Abell
catalog (sample ACO.A1, see Table~1 below) has a maximal spatial
density near the observer. The density decreases slowly up to a
distance $r\approx 375$~\Mpc; at this distance the spatial density is
approximately 75~\% of its local value.  At larger distances the
density rapidly decreases to zero at the formal upper limit of the
catalog that corresponds to a distance $r=405$~\Mpc.  It is
interesting to note that the spatial density of clusters located in
rich ($N_{cl} \ge 4$) and very rich ($N_{cl} \ge 8$) superclusters
follows the same trend, though at lower density levels, except at very
small distances where the density of clusters in very rich
superclusters drops to zero due to the absence of nearby very rich 
superclusters.  We see also that on distances up to $r \approx
200$~\Mpc\ almost all clusters have measured redshifts; the fraction
of measured redshifts slowly decreases to about 50~\% at $r=300$~\Mpc,
and to about 30~\% at $r=350$~\Mpc.

The spatial density of APM clusters shows a completely different
dependence on distance.  There are no APM clusters at small distances,
$r \le 50$~\Mpc, and almost no clusters in rich and very rich
superclusters at distances less than 150~\Mpc.  The overall density of
the sample increases approximately linearly with increasing distance
up to $r =r_{max}$, and thereafter drops suddenly to zero at
$r=350$~\Mpc\ for clusters in superclusters; there is a very sparse
extension of the distribution of isolated clusters up to a distance $r
\approx 450$~\Mpc.  At small distances   almost all clusters have
measured redshifts; with increasing distance the fraction of clusters
with measured redshifts continuously decreases; near the maximum of
the spatial density only 1 of 6 clusters has a measured redshift.

The selection function in Galactic latitude, $b$, is plotted in
Figure~4 where surface density of clusters is expressed in arbitrary
units. As argument we use $\sin b$, which is proportional to surface
area.  We do not distinguish between northern and southern Galactic
hemispheres of the Abell sample and plot a merged distribution.
Figure~4 shows that the surface density decreases almost linearly with
the decrease of $\sin b$.  This effect was noticed already by Einasto
\etal (1997b, hereafter E97b), and by E97d.  Conventionally the
selection effect in Galactic latitude is taken into account by cutting
the sample at some latitude ($|b| \approx 30^{\circ}$), and ignoring the
latitude dependence at high latitudes.  As we see, the latitude
dependence exists also at high latitudes, and cannot be ignored.
Figure~4 also shows that the difference between the latitude selection
of all clusters and clusters with measured redshifts is small.  A much
steeper dependence exists for clusters in rich and very rich
superclusters.  We shall discuss this difference in more detail in
Section 5.

\begin{figure*}[ht]
\vspace*{7cm}
\figcaption{Multiplicity functions for ACO and APM superclusters.  Solid
lines show the fraction of isolated clusters ($N_{cl}=1$) for
different neighborhood radii; dashed lines show the fraction of
clusters in superclusters of multiplicity $2 \le N_{cl} \le 31$; dot-dashed
lines indicate the fraction of clusters in extremely rich systems with
$N_{cl} \ge 32$.
}
\includegraphics{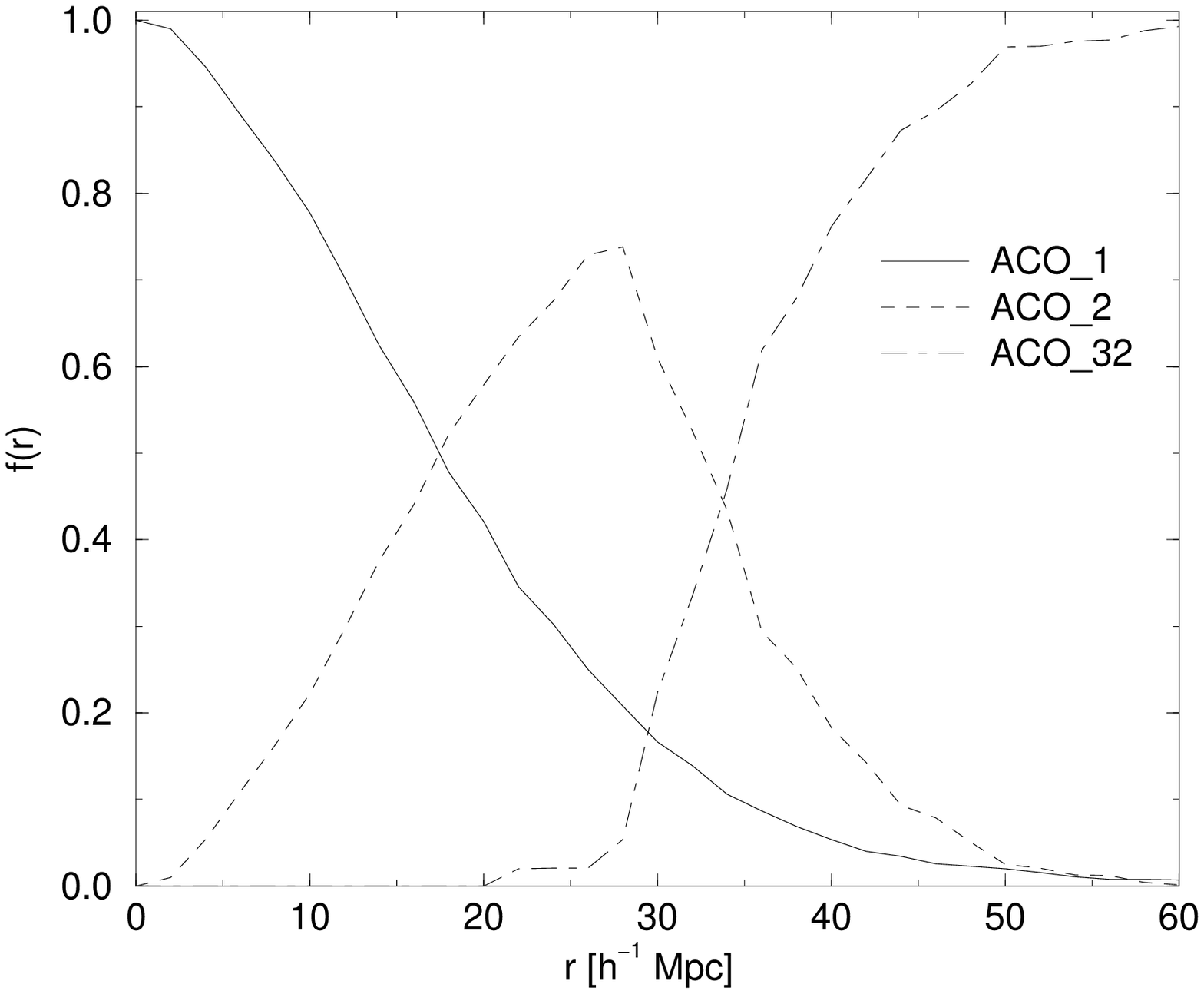} 
\includegraphics{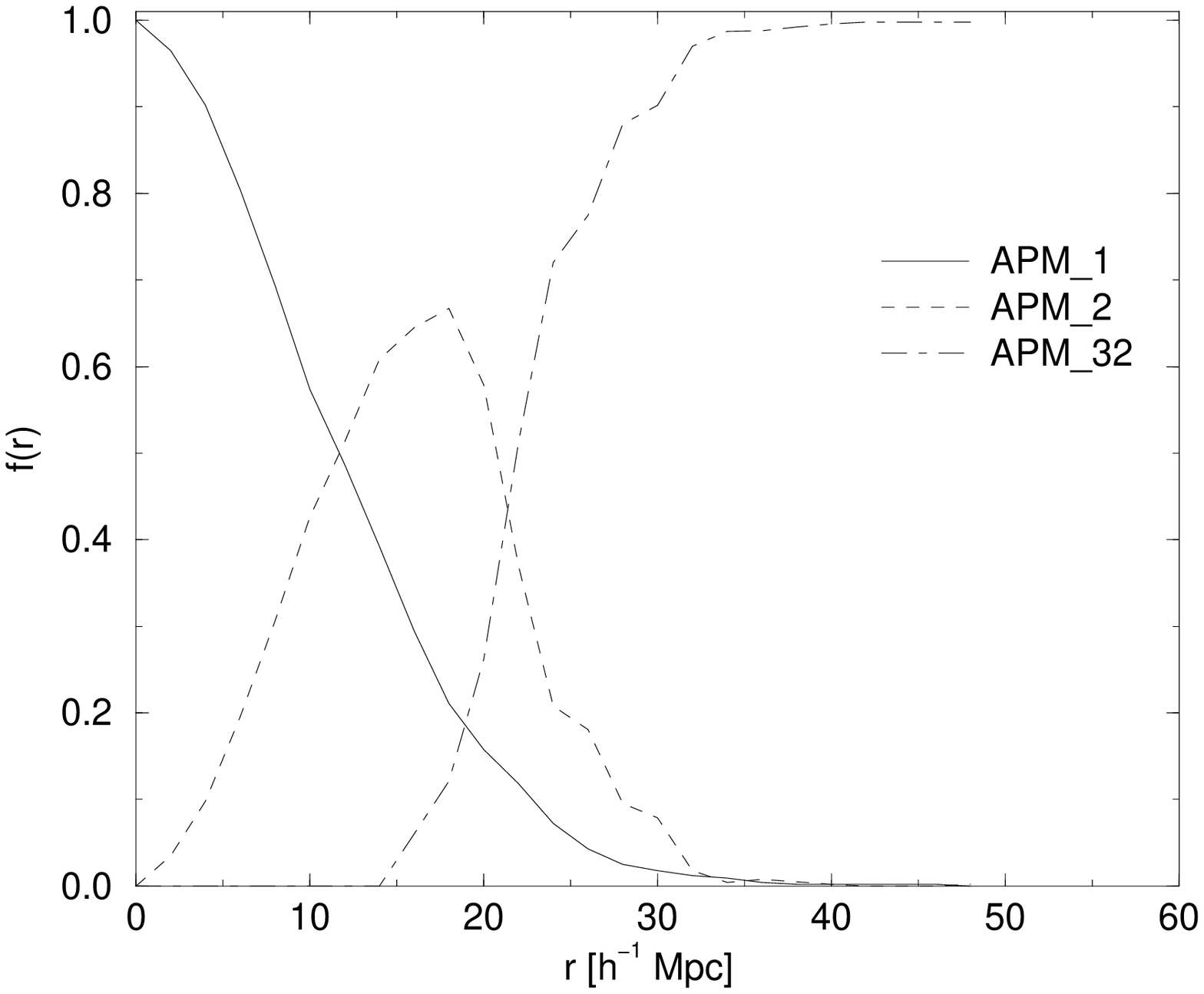}
\label{figure5}
\end{figure*}

The latitude dependence for APM clusters is shown in the right panel
of Figure~4.  The APM sample covers a restricted area around the
southern Galactic pole, thus the latitude dependence is steeper than
in the case of Abell clusters.  For comparison we plot the latitude
dependence for a model; it consists of 4000 points evenly distributed
on the sky between $-50^{\circ} \le RA \le 80^{\circ}$ and $-0.95 \le
\sin \delta \le -0.30$, covered by the APM sample.  Figure~4 shows
that the actual distribution of APM clusters has a steeper latitude
dependence than the model; in other words, the APM sample is subject
to latitude selection effects similar to that for the Abell sample.
We shall discuss selection effects in more detail in section 5.1.

Our previous studies of the correlation function and power spectrum
show that on large scales these functions are determined by clusters
located in high-density regions, i.e.\ in rich and very rich
superclusters (E97b, Einasto \etal 1997c, hereafter E97c).  Sparse
samples, where superclusters cannot be determined, have little or no
influence on these functions.  For this reason we shall use for
further study only cluster samples in a volume within a limiting radius
$r_{lim} = 350$~\Mpc, which corresponds to a limiting redshift of
approximately $z_{lim}=0.13$.  Beyond this distance the spatial
density of Abell clusters decreases rapidly.  We apply this distance
limit to Abell and APM samples. Data on samples used for further study
are given in Table~1.  These samples are designated as follows: the first 3
letters, ACO or APM, denote clusters from the respective catalogs;
the following capital letter denotes samples of all clusters, (``A''
including clusters with estimated redshifts), or clusters with
measured redshifts only (``R''); the following number is the minimum
richness of superclusters, $N_{cl}$: samples denoted by ``1'' include all
clusters (including isolated ones, i.e.  superclusters of richness
$N_{cl}=1$); samples designated by ``4''  are for clusters located in rich
superclusters which have at least 4 member clusters, $N_{cl} \ge 4$;
finally samples designated by ``8''  include very rich superclusters with at
least 8 members, $N_{cl}=8$.  $N$ denotes the number of clusters in
samples.  In Table~1 we give parameters of the selection functions,
which shall be described in Section~5.1.  Finally, we give in Table~1
also the correlation length, $r_0$.

\begin{table*}[ht]
\begin{center}
\caption[dummy]{Parameters of selection functions}
\begin{tabular}{lrrcccrcc}
\\
\hline
\hline
\\
Sample   & $N_{cl}$ & $N$  & $\sin b_0$ &$d_{0N}$&$d_{1N}$
& $d_{0S}$ & $d_{1S}$ & $r_0$ \\
\hline
\\
ACO.A1   & 1 & 1663  & 0.20 & 0.90 & 0.34 & 1.00 & 0.38 & 16 \\
ACO.A4   & 4 &  741  & 0.30 & 0.80 & 0.35 & 1.00 & 0.50 & 36  \\
ACO.A8   & 8 &  373  & 0.40 & 0.80 & 0.50 & 1.00 & 0.62 & 50  \\
\\
ACO.R1   & 1 & 1071  & 0.35 & 0.90 & 0.70 & 1.00 & 0.80 & 20 \\
ACO.R4   & 4 &  468  & 0.35 & 0.80 & 0.60 & 1.00 & 0.80 & 46 \\
ACO.R8   & 8 &  253  & 0.40 & 0.80 & 0.70 & 1.00 & 0.90 & 52 \\
\\
	&    &	     &	    &	   & $f_0$& $r_{min}$&$r_{max}$ & \\
\\
APM.A1   & 1 &  939  & 0.30 &      &  1.00& 50 & 310    & 13  \\
APM.A4   & 4 &  427  & 0.30 &      &  1.00& 100 & 310   & 33  \\
APM.A8   & 8 &  240  & 0.30 &      &  1.00& 150 & 310   & 49  \\
\\
APM.R1   & 1 &  356  & 0.30 &      &  0.70& 50 &  190    & 22  \\
APM.R4   & 4 &  159  & 0.30 &      &  0.70& 100 & 190    & 41  \\
APM.R8   & 8 &   95  & 0.30 &      &  0.70& 150 & 170    & 59  \\
\\
\hline
\label{tab:prop}
\end{tabular}
\end{center}
\end{table*}

\section{Clustering of Abell and APM clusters}

Galaxies and clusters of galaxies form large systems, called
superclusters.  We define superclusters as the largest relatively
isolated density enhancements in the Universe, which contain at least
one rich cluster of galaxies.  An example of a nearby supercluster
that contains only one rich cluster is the Local Supercluster.
Redshift data on individual galaxies become incomplete at large
distances, thus superclusters, as indicators of high-density regions
of the Universe, are usually found on the basis of only cluster data.

\subsection {Superclusters as defined by Abell and APM clusters}

The method to construct a supercluster catalog has been described in
detail by EETDA, E97d, and in Paper I.  To find high-density regions
from cluster data we use the friends-of-friends algorithm. Clusters
are assigned to superclusters using a certain neighborhood radius,
such that all clusters in the system have at least one neighbor at a
distance not exceeding this radius.  This neighborhood radius to
collect clusters to superclusters should be chosen in accordance with
the spatial density of the cluster sample as follows.  In Figure~5 we
show the fraction of clusters in superclusters of different
multiplicity for a wide range of neighborhood radii.  At small radii
all clusters are isolated, i.e. they belong to superclusters of
multiplicity $N_{cl}=1$. With increasing neighborhood radius some
clusters form superclusters of intermediate richness; $2 \le N_{cl}
\le 32$. At still larger radii (of $15 - 25$~\Mpc) extremely large
superclusters with multiplicity $N_{cl} \ge 32$ start to form. By
further increasing the neighborhood radius to $\approx 30$~\Mpc,
superclusters begin to join into huge conglomerates; finally, at the
``percolation radius'' of $\approx 45$~\Mpc\ all clusters percolate
and form a single system penetrating the whole space under study.  We
want to select superclusters as large as possible, but which are still
isolated systems.  To obtain superclusters with these properties we
must choose a neighborhood radius well below the percolation radius.
The appropriate neighborhood radius is the radius which corresponds to
the maximum of the fraction of clusters in systems of intermediate
richness.  At this radius very large systems just start to form, as
seen from Figure~5 (see also EETDA, E97d and Paper I).

The spatial density of APM clusters is about three times that of Abell
clusters (section 5.2, see also Dalton \etal 1992, Croft \etal 1997).
This property is reflected in the multiplicity function: these
functions are similar for the APM and Abell clusters, but for the APM
sample they are shifted to smaller neighborhood radii with
respect to those for the Abell sample. For the Abell cluster sample 
we chose a neighborhood radius of 24~\Mpc\ to derive a supercluster
catalog, while the APM supercluster catalog was based on a
radius of 16~\Mpc.  Both values were chosen so as to maximize the
fraction of systems of intermediate richness.  The most recent Abell
supercluster catalog (as presented in Paper~I) is based on all 
clusters up to a limiting redshift of $z_{lim}=0.13$.  The supercluster
catalog based on APM clusters is given in the Appendix of the present
paper.

\subsection{The distribution of clusters in rich  superclusters}

Now we shall compare the spatial distribution of superclusters based
on the Abell and APM cluster samples.  We restrict our comparison to
the area of the sky covered by the APM sample.  We use supercluster
catalogs constructed for all clusters (including clusters with
estimated redshifts). We show the distribution of clusters only in
superclusters of richness 4 and higher. This has two reasons.  First,
the skeleton of the supercluster-void network is essentially
determined by rich superclusters; poorer systems only add some fine
details to the structure (E97d).  Redshift errors do not affect the 
sky positions of the clusters but have an effect on their radial 
distance only.  Redshift errors have the same
effect for superclusters as the velocity dispersion of galaxies in
clusters: they elongate the superclusters in the radial direction.
This effect may destroy real superclusters, if the separation between
individual clusters in the radial direction, due to errors, is too
large.  Using rich and very rich superclusters we have a higher
probability to select real superclusters since it is very unlikely
that errors bring isolated clusters together to form compact rich
systems.

\begin{figure*}[ht]
\vspace*{13cm}
\figcaption{The distribution of Abell clusters (left-hand panels) and
APM clusters (right-hand panels) in rich and very rich superclusters
in equatorial coordinates (RA is given in radians, in vertical axis we
plot $\sin \delta$, both for equinox 1950).  Large
symbols mark very rich superclusters ($N_{cl} \ge 8$), small
symbols clusters in 
superclusters of intermediate richness ($4 \le N_{cl} < 8$);
open symbols are for clusters with estimated, filled symbols for
measured redshifts. The upper panels plot clusters in distance interval
$250 \le r < 350$~\Mpc, while the lower panels in interval $150 \le r <
250$~\Mpc.  Centers of very rich superclusters determined from the
APM.R8 cluster sample are shown by extra-large open circles. }
\includegraphics{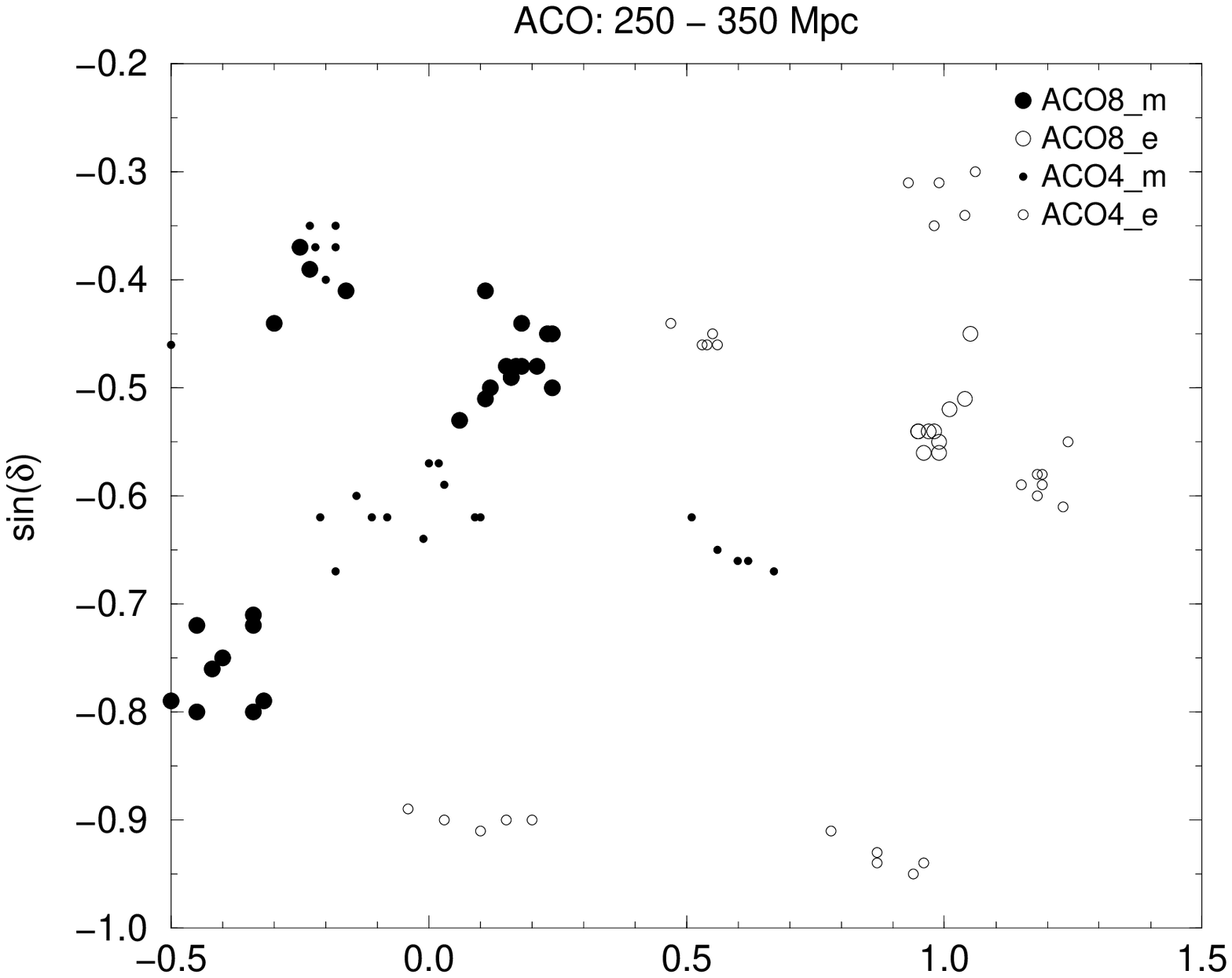} 
\includegraphics{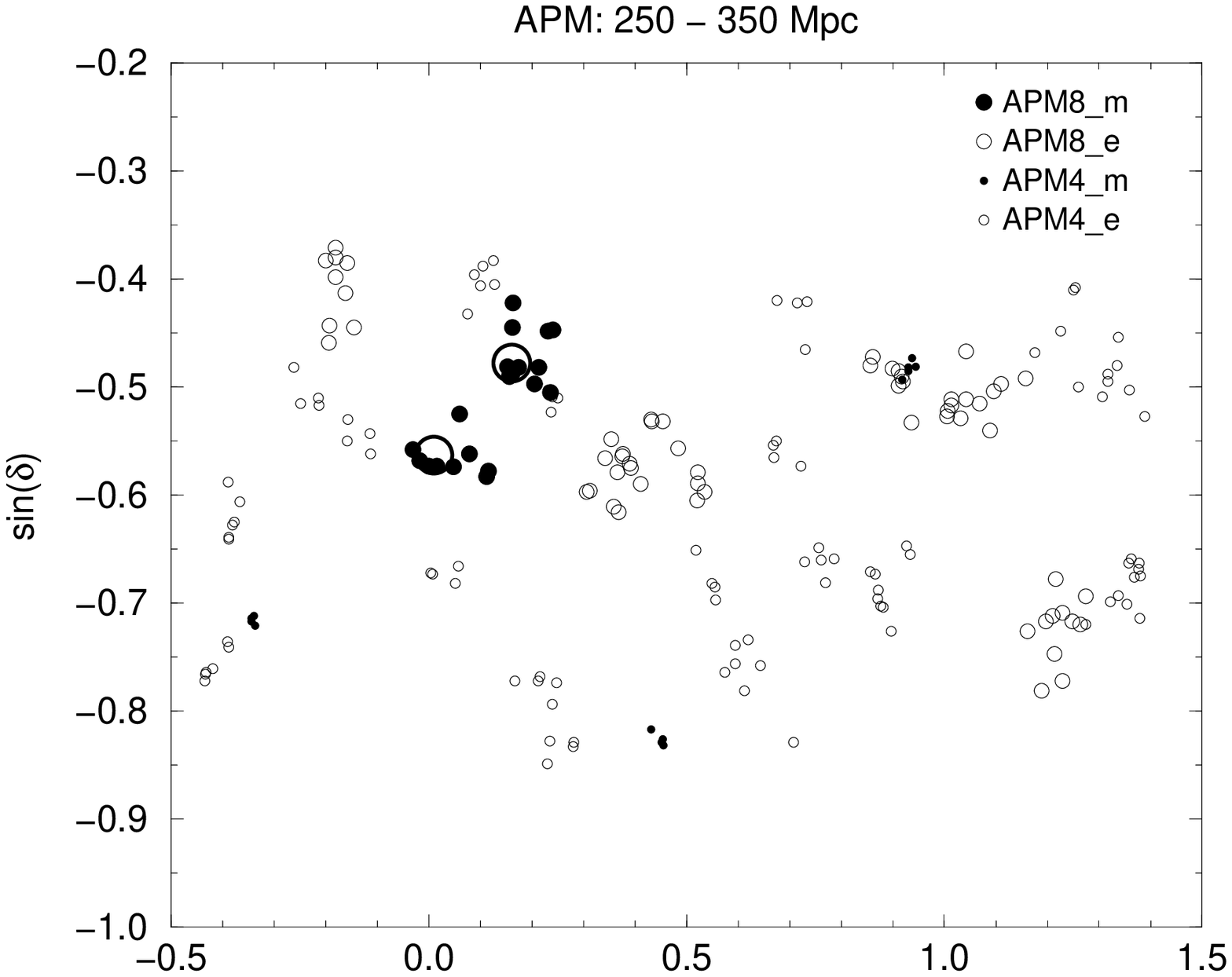} 
\includegraphics{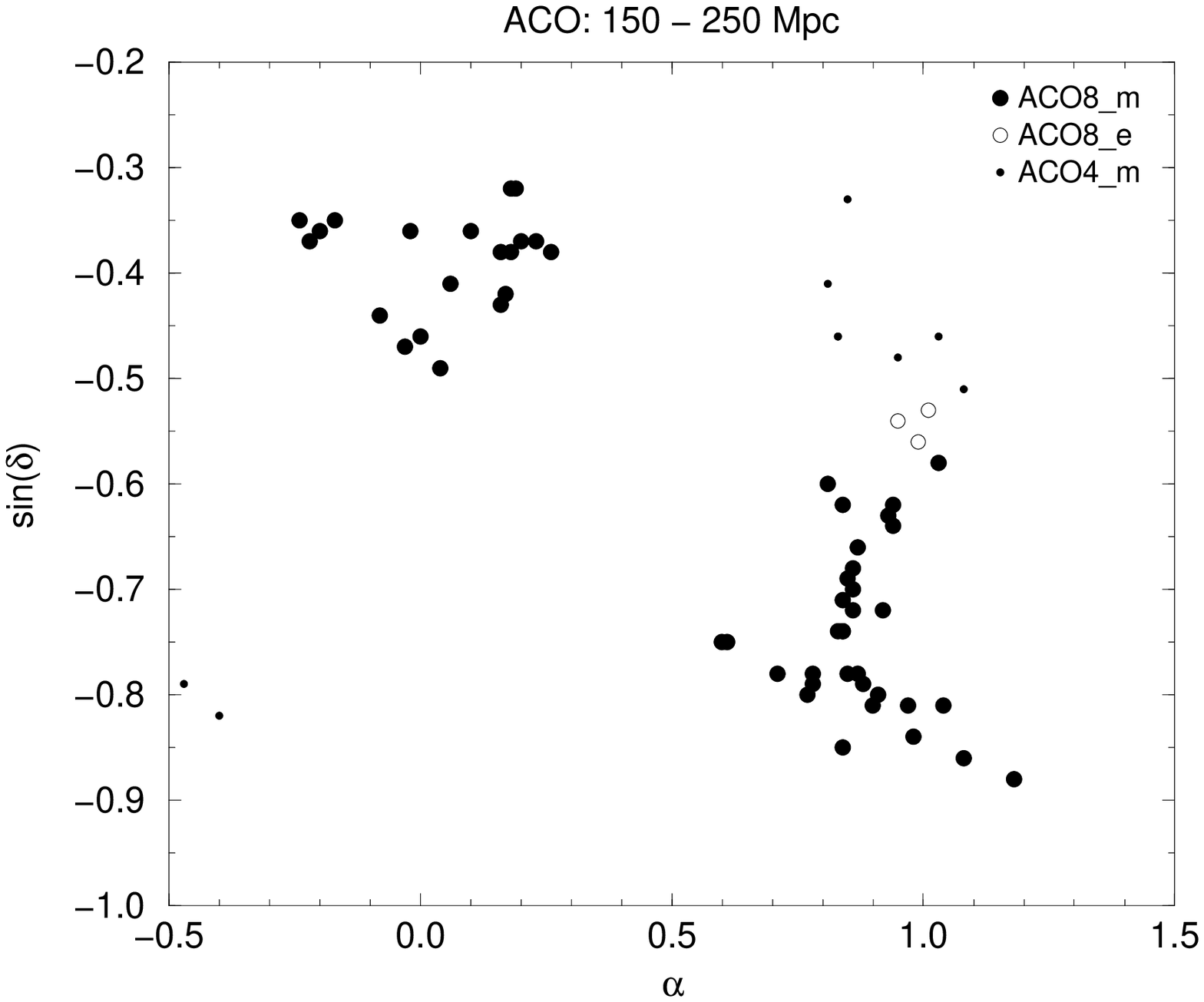} 
\includegraphics{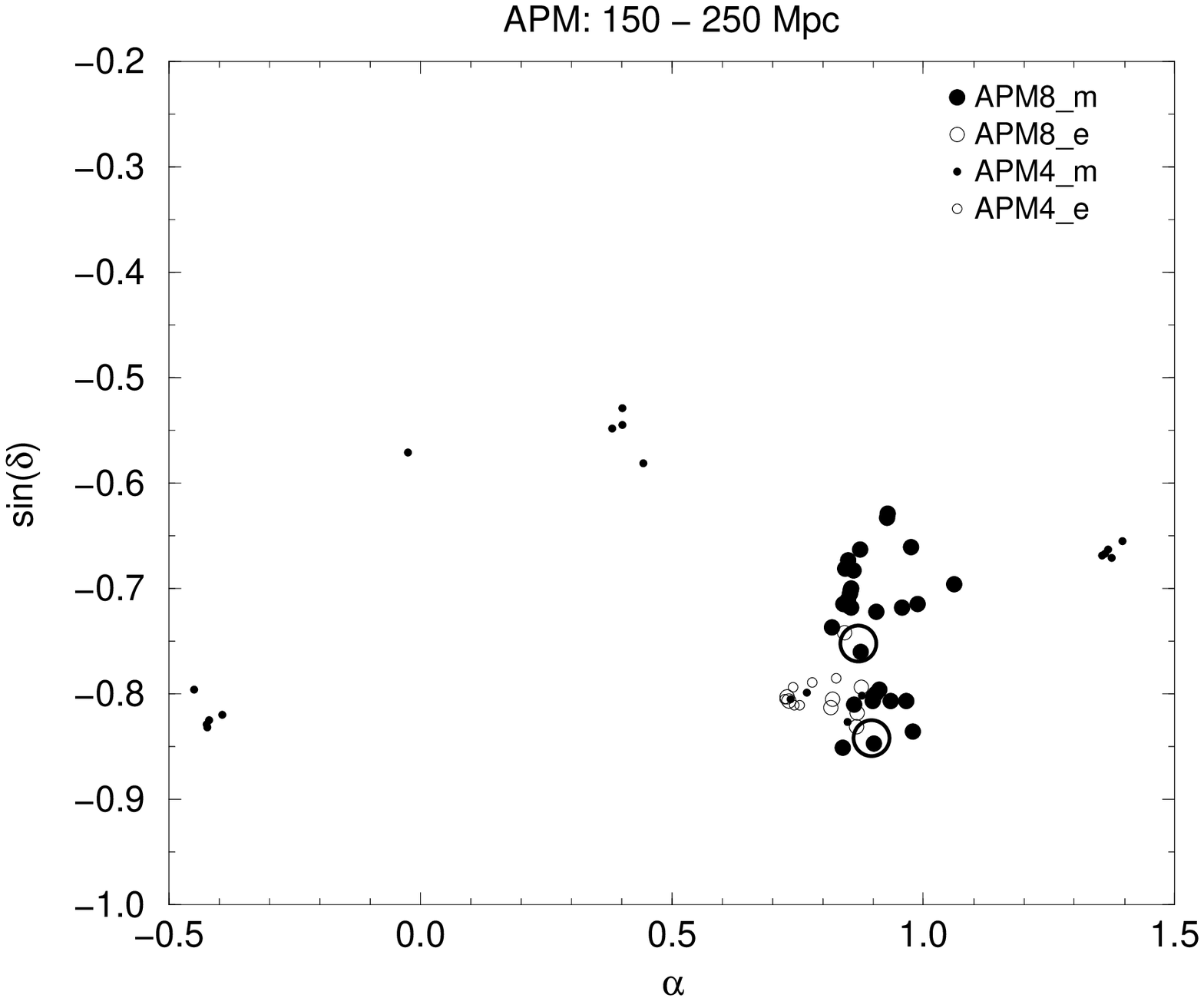}
\label{figure6}
\end{figure*}

Figure~3 shows that in this region of the sky the distribution of
clusters has two maxima, located at distances $r \approx 200$ and
300~\Mpc\ from us.  The APM supercluster catalog contains no rich
superclusters at distances less than 150~\Mpc, so we plot in Figure~6
clusters in rich and very rich superclusters in two layers, $150 \le r
< 250$~\Mpc, and $250 \le r < 350$~\Mpc, which correspond to maxima of
the cluster sample density.  The richest superclusters in the nearer
sheet (lower panels in Figure~6) are the Pisces-Cetus, Grus-Indus,
Horologium-Reticulum (see also Park \& Lee 1998) superclusters located
at RA $\approx 0.1\,rad, ~\sin(DEC) \approx -0.4$; RA\,$\approx -0.6,
~\sin(DEC) \approx -0.8$, and RA\,$\approx 0.9, ~\sin(DEC) \approx
-0.8$, respectively.  The richest superclusters in the farther sheet
(upper panels) are the Sculptor supercluster at RA$ \approx 0.1, ~\sin
\delta \approx -0.5$, and the Caelum and Fornax-Eridanus supercluster
complex at RA $ \approx 1.1, ~\sin \delta \approx -0.6$.  Very rich
superclusters are seen in both (the Abell and APM) supercluster
catalogs based on all clusters, although individual clusters differ.

Extra-large open circles in Figure~6 mark the centers of very rich
superclusters as determined from positions of their member clusters in
the APM.R8 supercluster catalog.  In the nearer sheet only the
Horologium-Reticulum supercluster complex has a sufficient number of
members with measured redshifts, and is split into two subsystems,
separated by 25~\Mpc\ in 3-D space.  In the farther sheet we see the
Sculptor supercluster which is also split into two subsystems
(separated by 50~\Mpc\ in 3-D space, see Figure~9).  We conclude that
the whole APM supercluster catalog, derived on the basis of measured
redshifts, is dominated by two very rich supercluster complexes, the
Horologium-Reticulum and the Sculptor superclusters.  Other rich
superclusters in the APM sample have only a few, if any, clusters with
measured redshifts.  These dominant rich superclusters surround the
Sculptor void, one of the largest voids known with a diameter of about
185~\Mpc. On the borders of this void there are several APM clusters
forming a cluster filament. The spatial density of this filament is,
however, lower than the minimum density below which for clarity we
omit clusters in low-density environments in Figure~6.

The sample of all APM clusters (APM.A1) reaches its maximal spatial
density at a distance $r \approx 300$~\Mpc\ (see Figure~3).  It is
interesting to note that clusters which belong to rich and very rich
superclusters of the APM supercluster catalog (upper right-hand
panel of Figure~6) form a quasi-regular pattern on the sky with a
period of about 0.4 radians along the RA axis. At a distance of
300~\Mpc\ this period corresponds to a linear separation of $\approx
120$~\Mpc.  As the sky positions of superclusters are not affected by
distance errors, this period cannot be much in error.  We see that the
whole APM cluster catalog gives a visual impression of a
supercluster-void network with a characteristic scale of $\approx
120$~\Mpc\ between high-density regions, similar to the scale found on
the basis of Abell clusters (EETDA, E97d).

\begin{figure*}[ht]
\vspace*{7cm}
\figcaption{The correlation functions of clusters in very rich
superclusters for Abell (left hand panel) and APM samples (right-hand
panel).  Solid lines show correlation functions of all clusters
located in very rich superclusters (samples ACO.A8 and APM.A8); dots
with error bars designate correlation functions on the basis of
cluster samples with measured redshifts (samples ACO.R8 and APM.R8).
In the right-hand panel the dashed line marks the correlation function
of ACO.A8 clusters for comparison.
}
\includegraphics{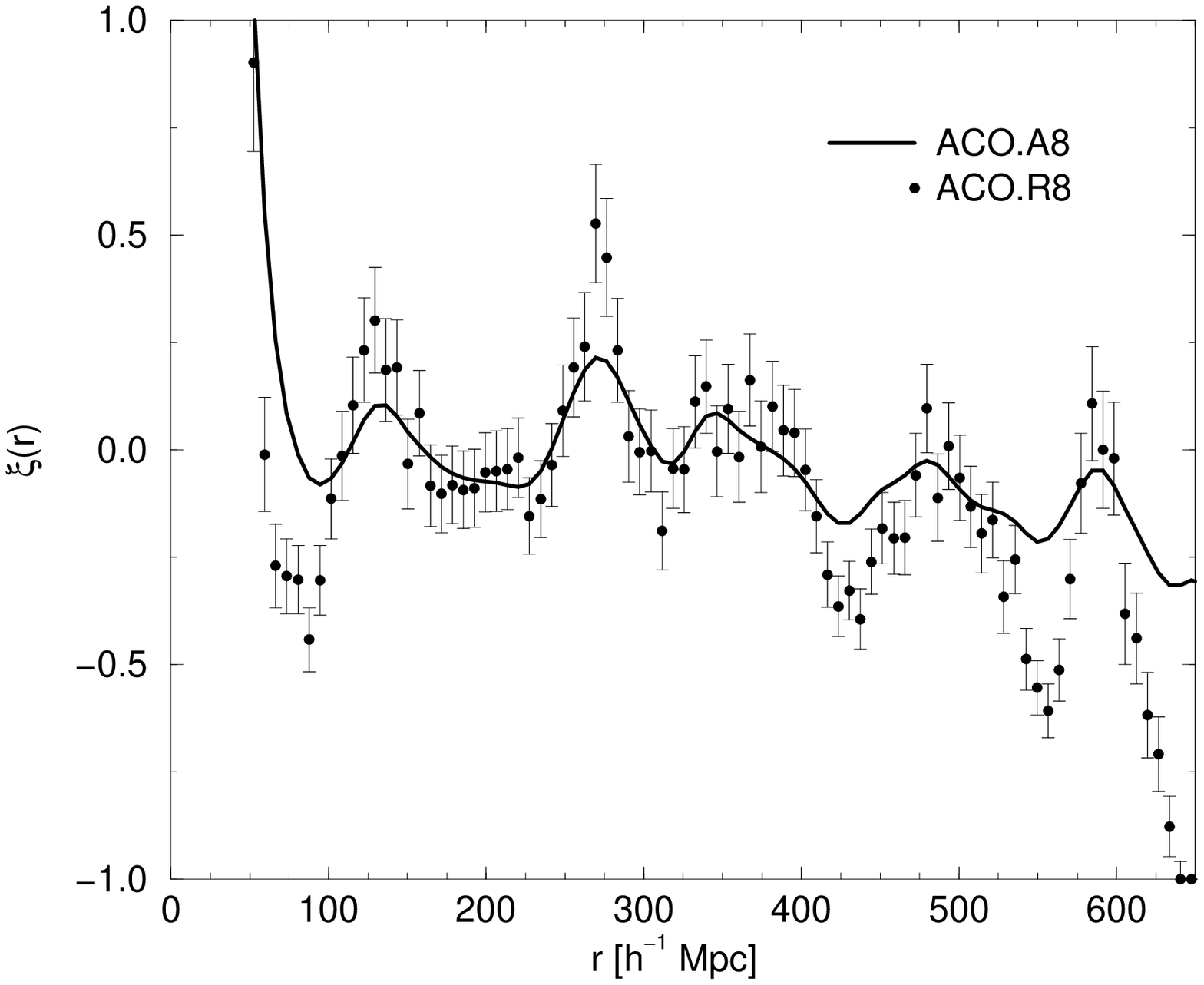} 
\includegraphics{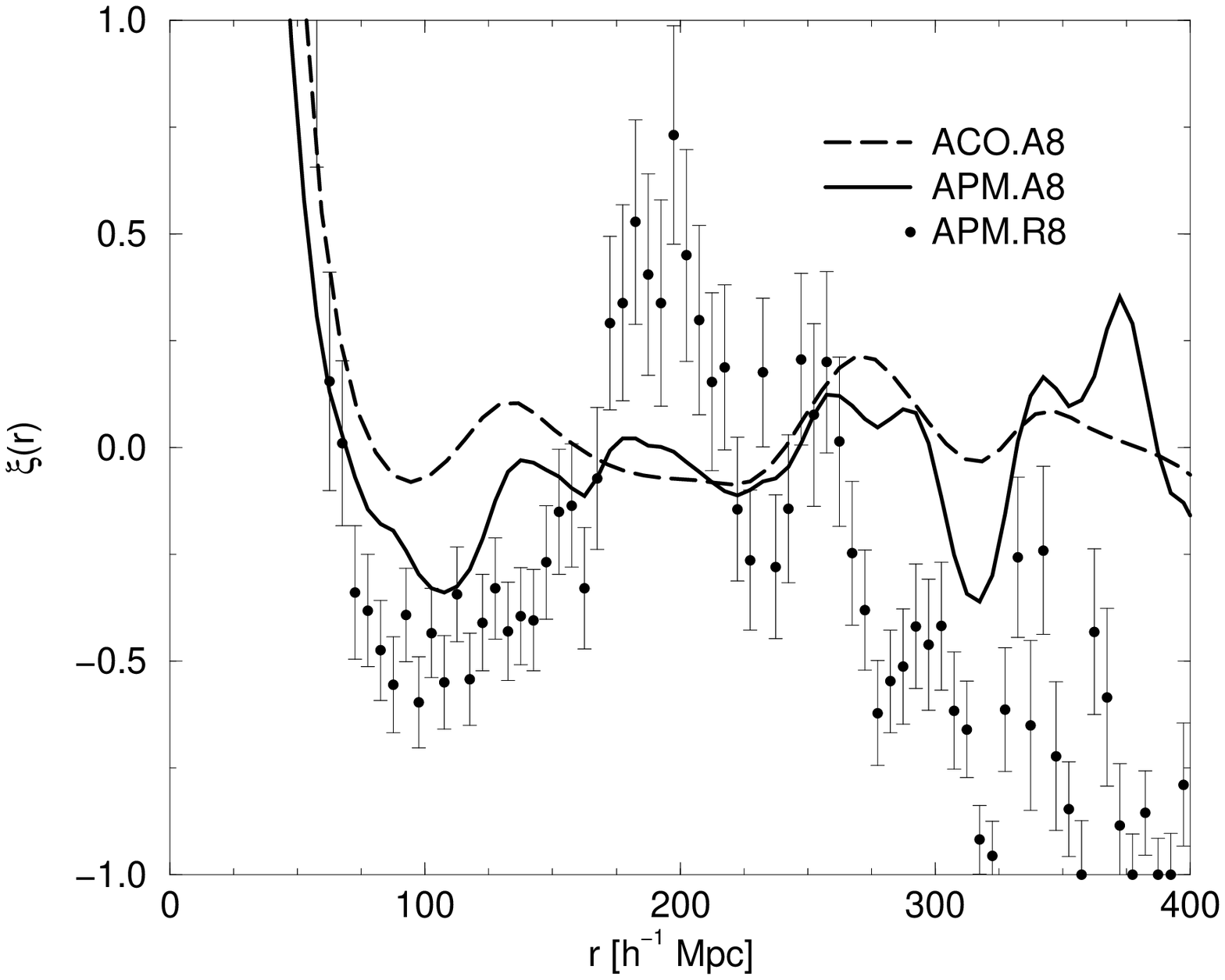}
\label{figure7}
\end{figure*}

\section{Quantitative measures of clustering of  Abell and
APM clusters}

\subsection{Correlation functions}

The correlation functions were calculated using the classical method:
\begin{equation}
\xi(r)={\langle DD(r)\rangle  \over \langle RR(r)\rangle }{n_R^2
\over n^2} -1, 
\label{eq1}
\end{equation}
where $\langle DD(r)\rangle $ is the number of pairs of clusters in
the range of separations $r\pm dr/2$, $dr$ is the bin size, $\langle
RR(r)\rangle $ is the respective number of pairs in a Poisson sample
of points, $n$ and $n_R$ are the mean number densities of clusters in
respective samples, and brackets $\langle \dots \rangle $ mean
ensemble average. The summation is over the whole volume under study,
and it is assumed that the cluster and Poisson samples have identical
shape, volume and selection functions. No weights are used; the
selection effects are applied to Poisson samples and all real and
simulated points are treated with equal weights.  The mean error
(determined by cosmic variance of samples) was calculated using the
equation:
\begin{equation}
\sigma_{\xi c}={b \over \sqrt{N}}, 
\label{eq2}
\end{equation}
where $N$ is the number of clusters in the sample, and $b=1.5$ is a
parameter that describes the character of the large-scale distribution
of objects studied (E97b, E97c).

Correlation functions were calculated for all Abell and APM cluster
subsamples listed in Table~1. These functions are determined basically
by the distribution of clusters which belong to rich and very rich
superclusters (E99a). To show the properties of the distribution of
high-density regions we plot in Figure~7 correlation functions of
clusters in very rich superclusters, samples ACO.A8, ACO.R8, APM.A8
and APM.R8.  To generate comparison Poisson samples we used selection
functions described in section 5.1 below, with parameters given in
Table~1.  For samples ACO.R8 and APM.R8 we plot values of the
correlation function with error bars, for samples ACO.A8 and APM.A8 we
show correlation functions smoothed with Gaussian window of dispersion
10~\Mpc.

Abell clusters occupy a double-conical volume with full depth of
700~\Mpc, thus it is possible to calculate the correlation function
for large separations.  The APM sample is defined in a smaller volume,
and the correlation function can be found for smaller separations.  We
see that correlation functions, derived for Abell clusters in very
rich superclusters, are oscillating. We can recognize 5 secondary
maxima and 6 minima. The mean separation of maxima and of minima is
$116 \pm 21$~\Mpc.  The differences between correlation functions
derived for all clusters and for clusters with measured redshifts
(samples ACO.A8 and ACO.R8, respectively) are small.

The correlation function of APM clusters has a more complicated
behavior.  If we use all clusters (sample APM.A8), then the first and
the second secondary maxima have locations close the locations of
respective maxima found for Abell cluster samples.  Similarly we can
identify the first and the third minima with minima in the Abell
cluster correlation function.  But instead of the second minimum near
separations of $r \approx 200$~\Mpc, the APM sample has a maximum at
this separation, not present in the Abell sample.  If we use only
clusters with measured redshifts (sample APM.R8), then the first
secondary maximum of the correlation function at $r \approx 130$~\Mpc\
disappears, and the peculiar maximum at $r \approx 185$~\Mpc\ has an
enhanced amplitude.

The reason for such peculiar behavior can be understood when we
consider the distribution of very rich superclusters in the APM
samples.  The sample with all clusters (APM.A8) is dominated by
numerous very rich superclusters located in the more distant shell
(see right-hand upper panel of Figure~6).  These superclusters are
distributed fairly regularly and form a supercluster-void network with
a step around 120~\Mpc\ (see previous Section).  If we use the sample
with measured redshifts only (APM.R8) instead, then the number of very
rich superclusters in the sample decreases; the sample is dominated by
two very rich superclusters that border the Sculptor void, one of the
largest voids known (see the extra-large circles in the right panels
of Figure~6 and a jump at 200~\Mpc\ of distance distribution of APM.R8
clusters in Figure~9). The secondary maximum of the correlation
function of the sample APM.R8 is determined by mutual separations of
clusters belonging to these superclusters.  We shall discuss this
behavior of the APM sample below.

We also calculated the correlation length (i.e. the separation at
which the correlation function equals unity), $r_0$, for all samples.
The results, given in Table~1, show that there are only minor
differences between Abell and APM cluster samples.  In both cases the
correlation length, determined for all clusters, is $r_0 = 18 \pm
3$~\Mpc; for clusters in rich superclusters it is $r_0 = 39 \pm
5$~\Mpc, and for clusters in very rich superclusters, $r_0 = 52 \pm
7$~\Mpc\ (the mean and scatter are determined from all Abell and APM
samples with respective $N_{cl}$).  Similar values have been found by
E97b.

\begin{figure*}[ht]
\vspace*{7cm}
\figcaption{Power spectra of ACO (left) and APM clusters of galaxies
(right panel).  Different symbols are used for spectra based on all
clusters and for spectra derived for samples of clusters with only
measured redshifts.  The error corridor shown for spectra based on
samples ACO.R1 and APM.R1 was calculated from the $3\sigma$ error of
the correlation function.  For comparison we show also spectra found
by E97a and R98 for Abell clusters, and by T98 for APM clusters.}
\includegraphics{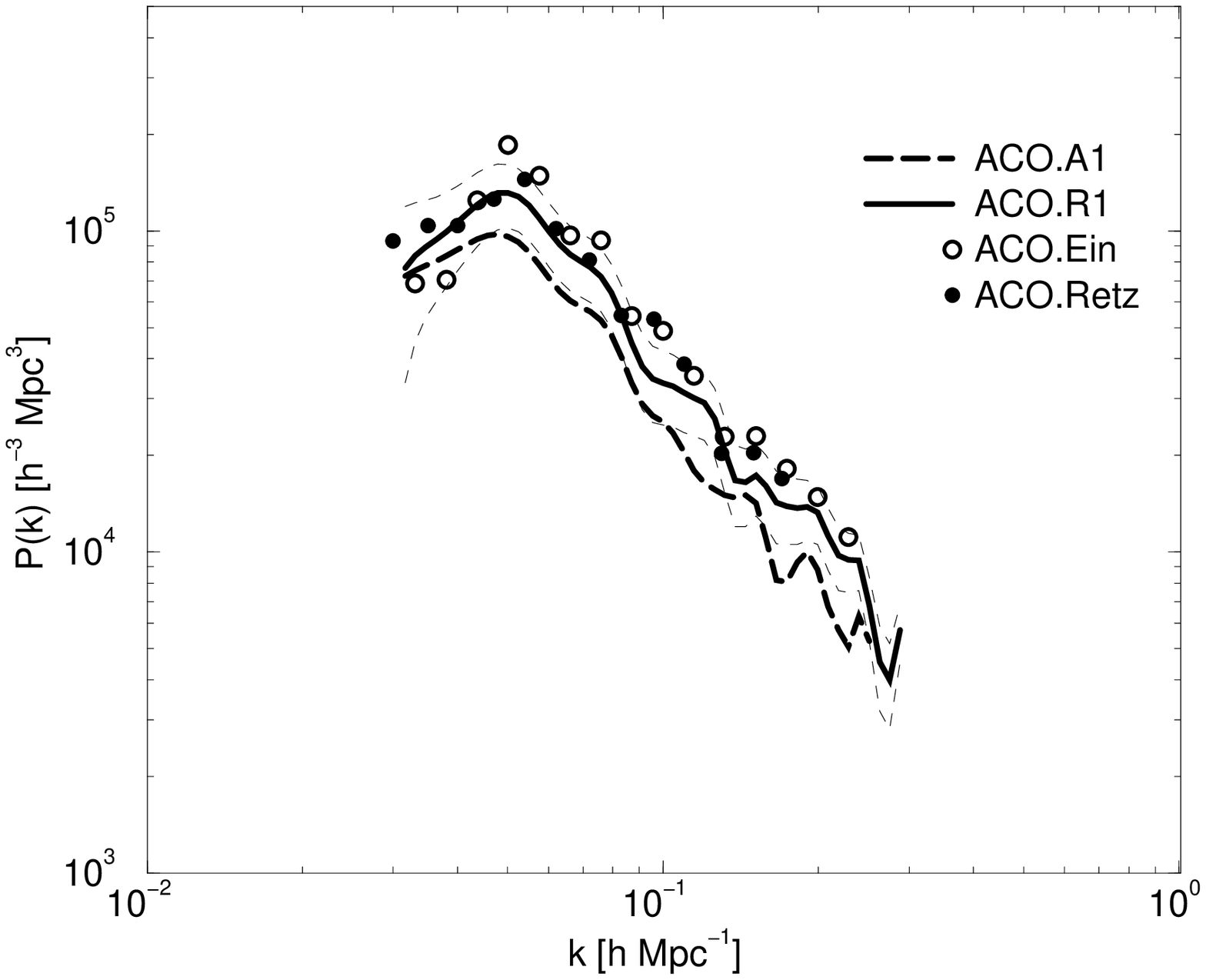} 
\includegraphics{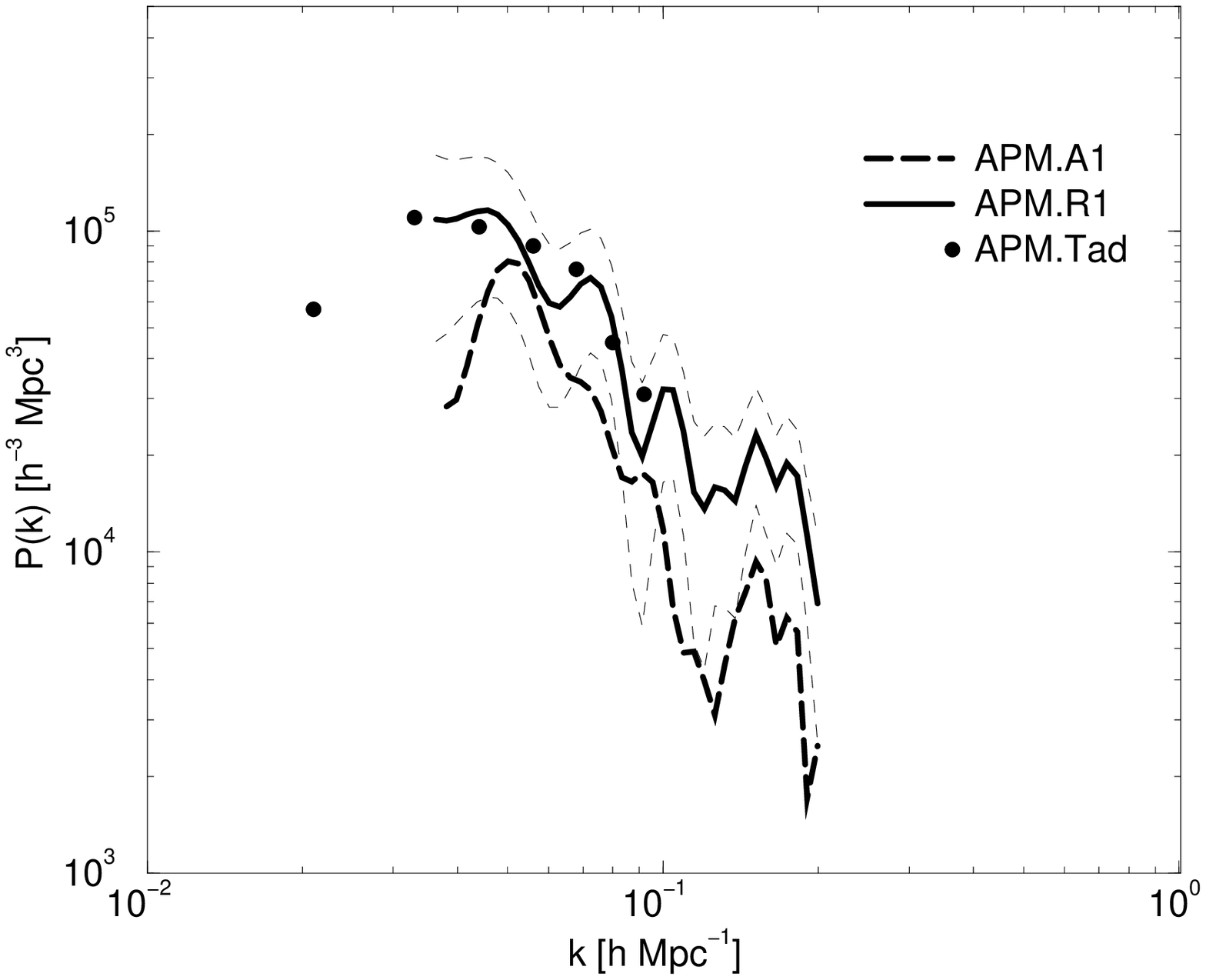}
\label{figure8}
\end{figure*}

\subsection {Power spectra}

The power spectrum, $P(k)$, is  the  the Fourier transform of the
correlation function, $\xi(r)$, and vice versa:
\begin{equation}
P(k)=4\pi\int_0^{\infty}{\xi(r) r^2 {\sin kr \over kr} dr};
\label{eq3}
\end{equation}
\begin{equation}
\xi(r)={1 \over 2\pi^2}\int_0^{\infty}{ P(k) k^2 {\sin kr \over kr} dk}.
\label{eq4}
\end{equation}
here the wavenumber $k$ is measured in units of $h~ {\rm Mpc^{-1}}$,
and is related to the wavelength by $\lambda=2\pi/k$.  Practical
procedures to calculate integrals (\ref{eq3}) and (\ref{eq4}) are
given by Press \etal (1992).  The Fourier transform yields accurate
results only if the correlation function (or the power spectrum) has
no errors, and if the power spectrum and the correlation function are
given for the whole $k$ or $r$ space, respectively.  These conditions
are not fulfilled in the case of the real samples.  The power spectrum
is, by definition, a non-negative quantity.  Integral (\ref{eq4})
yields for all non-negative power spectra a physically reasonable
correlation function, but not all correlation functions submit
non-negative results when one is calculating the power spectrum using
formula (\ref{eq3}).  For this reason random and systematic errors of
the correlation function can result non-physical values for the power
spectrum.

The most serious systematic error introduced by the cluster
correlation function is due to the peculiar shape of cluster samples.
The correlation function of clusters in superclusters has an
oscillatory behavior: at large separations it contains regularly
spaced secondary maxima and minima.  Maxima correspond to mutual
separations between rich superclusters, and minima to mean separations
of voids.  If the sample occupies fully a large truly 3-D volume, then
the amplitude of oscillations of the correlation function decreases
almost exponentially, since on large separations rich superclusters
located in different directions reduce the maxima and minima of the
correlation function (see E97c for a discussion).  If the sample
volume is not fully 3-dimensional, then the cancellation is less
effective.  In the extreme case, if the sample is given in a
cylindrical (essentially one-dimensional) volume, and if superclusters
are regularly spaced along the axis of the sample, then the amplitude
of oscillations of the correlation function does not decrease with
increasing separation of superclusters.  If we take a double-conical
volume, as is the actual Abell sample volume that contains no clusters
near the Galactic equator, then the cancellation of maxima and minima
is partial.  In a sample of such elongated form the amplitude of
oscillations of the correlation function decreases with the increase
of the separation more slowly than in the case of a fully 3-D sample.

To correct for such an effect we have to suppress the amplitude of
oscillations of the correlation function on large separations,
preserving the shape of the function on small separations.  Our
experience has shown that an appropriate correction can be achieved if
we multiply the observed correlation function by a factor $\exp
(-r/r_c)^2$, here $r_c$ is a parameter.  We have checked the recovery
procedure using correlation functions found via (\ref{eq4}) from
analytical power spectra, distorted to increase the amplitude of
oscillations, and thereafter corrected according to the procedure
outlined above.  This trial has shown that this procedure recovers the
actual shape of the power spectrum rather closely. By a
trial-and-error procedure we have found that a value $r_c=150$~\Mpc\
yields best recovery of analytical and observed correlation functions.

\begin{figure*}[ht]
\vspace*{7cm}
\figcaption{Cumulative distribution of distances between centers of
superclusters for Abell and APM very rich superclusters.
Only Abell superclusters in the APM window were considered.
}
\includegraphics{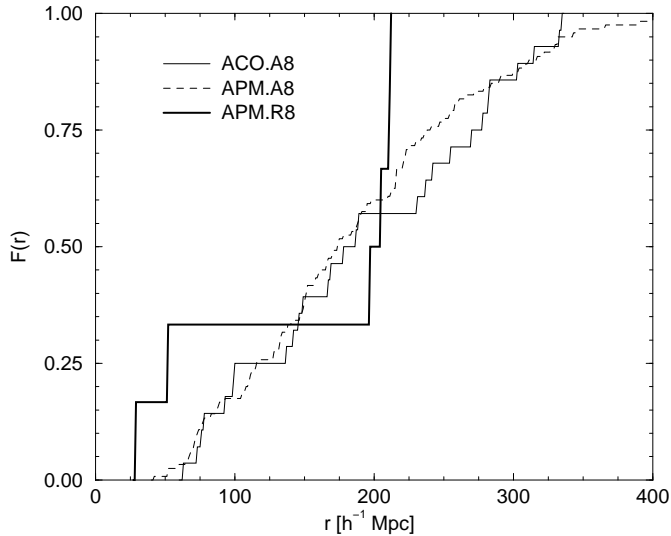} 
\label{figure9}
\end{figure*}

We have investigated also the influence of random errors on the
correlation function.  We added random noise to the distorted
correlation function; this noisy correlation function was reduced
using the procedure given above, and Fourier-transformed to obtain the
recovered power spectrum. This test has shown that purely random
errors have little influence.  Actually random, systematic (due to
sample shape) and cosmic (deviations of samples found for limited
volume from the underlying large parent sample) errors distort
observed correlation functions in a more complicated way. Thus the
recovered power spectrum may still have some further peculiarities.

The power spectra of ACO and APM cluster samples found on the basis of
their correlation functions are shown in Figure~8.  Here
we have used correlations functions of all clusters and of clusters
with measured redshifts only, samples ACO.A1, ACO.R1, APM.A1, and
APM.R1.  Our experience has shown that our recovery procedure yields
for real cluster samples meaningful results only in the wavenumber
interval $0.03-0.04 \leq k \leq 0.2-0.3$~\hmpc.  On smaller scales the
power spectrum is noisy and contains sections with negative values, on
larger scales the error corridor found on the basis of the error
corridor of the correlation function becomes very large.  For both
cluster samples the amplitude of the power spectrum for all clusters
is lower than for clusters with measured redshifts.  This is probably
due to observational errors which smooth the structure and reduce the
amplitude of density perturbations.  The shape of the power spectra
for samples ACO.A1 and ACO.R1 is very similar.  In contrast, the
position of the maximum of the power spectrum of samples APM.A1 and
APM.R1 is different: for APM.A1 it lies at $k\approx 0.05$~\hmpc,
similar to the maximum of Abell cluster samples, whereas for APM.R1 it
lies at $k\approx 0.04$~\hmpc.  This difference is due to differences
in correlation functions, compare with Figure~7.

\subsection {The distribution of distances between superclusters}

So far we have characterized the distribution of clusters using the
correlation function and its Fourier transform, the power spectrum.
An additional measure is the distribution of distances between
superclusters. While correlation functions and power spectra were
calculated on the basis of cluster samples, we now consider the
distribution of centers of superclusters.  Our main goal is to find
the reason why the secondary maximum of the correlation function for
the sample APM.R8 is shifted towards larger separations compared to
results obtained with Abell cluster and APM.A8 samples.  As these
functions are basically determined by the distributions of
high-density regions, we shall derive the distribution of mutual
distances of very rich superclusters with at least 8 members.  For
this we use the samples APM.A8 and APM.R8.  For comparison we also
derive the distribution of distances between Abell superclusters
(ACO.A8) located in the region of sky covered by the APM cluster
sample.  Integrated distributions of distances are shown in
Figure~9.

We see that there are practically no differences between the
distribution of distances of Abell and APM superclusters, at least if
we use superclusters based on samples including all clusters.  This
result confirms our earlier findings, based on the correlation
function, that Abell and APM samples of all clusters have similar
clustering properties.  In contrast, the distribution of distances
between rich superclusters, found for APM clusters with measured
redshifts, is very different.  The APM.R8 supercluster sample contains
only two supercluster complexes, the Horologium-Reticulum supercluster
and the Sculptor supercluster located on opposite sides of the
Sculptor void (Figure~3).   By chance this is the largest
cluster-defined void in our vicinity.  These very rich systems have
the strongest effect on the correlation function of APM clusters with
measured redshifts.  The distances between these superclusters
characterize the size of the Sculptor void between them, 185~\Mpc\
(the distances between supercluster centers are slightly larger, about
200~\Mpc).  As we have shown above, the correlation function of APM
clusters with measured redshifts has its secondary maximum exactly at
the same distances.  Thus, this maximum in the correlation function is
due to the mutual separation of these two superclusters.  For
comparison: the new Abell cluster sample contains 31 very rich
superclusters (see Table A1 in Paper I) and 16 voids cataloged by
EETDA and surrounded by rich superclusters.  To summarize: the Abell
cluster sample (and partly also the APM sample of all clusters)
characterizes mean properties of the supercluster-void network,
whereas the APM sample with measured redshifts reflects properties of
an extremely large deviation from the mean.

\section{Discussion}

In this section we discuss differences observed between two catalogs
of clusters, and between statistics derived from these catalogs in
more detail.  We shall concentrate on two aspects of the problem: the
selection function, and the correlation function.  Thereafter we shall
compare both cluster samples for their capabilities in tracing the
supercluster-void network.

\subsection{Selection effects in Abell and APM cluster samples}

Suppose that we study a certain volume in the Universe and have in
this volume a parent sample of clusters of galaxies.  Let
$N_{true}(V,{\cal R})$ be the actual number of clusters in this
volume, $V$, with a true limiting number of galaxies in the cluster,
$\cal R$, varying from cluster to cluster.  Galaxies are chosen within
a projected distance of one Abell radius (1.5~\Mpc) around the cluster
center, and no more than 2 magnitudes fainter than the 3rd-brightest
galaxy.  This working definition was used by Abell (1958) and ACO. The
cluster richness was then determined from the resulting number of
galaxies counted in a circle of angular radius based on an assumed
cluster distance, which was estimated from the apparent magnitude of
the 10th-brightest galaxy.  All quantities used in the cluster
definition are subject to random and systematic errors; thus the
observed number of clusters, $N_{obs}(V,{\cal R})$, differs from the
true one. In general, the numbers of observed and true clusters of
galaxies are related as follows:
\begin{equation}
N_{obs}(V,{\cal R})= \Phi(V,{\cal R}) N_{true}(V,{\cal R}),
\label{eq5}
\end{equation}
where
\begin{equation}
\Phi(V,{\cal R})=\prod \varphi_i(V,{\cal R}),
\label{eq6}
\end{equation}
and $\varphi_i(V,{\cal R})$ are probabilities to detect a cluster of
richness $\cal R$ in the volume $V$, given the various selection
effects, and the errors in the parameters used in the cluster
definition.  Here we assume that selection effects and errors $i$ are
independent of each other.  Most selection effects decrease the number
of observed clusters, i.e.\ the respective probabilities are
$\varphi_i \leq 1$.  Errors in parameters of cluster definition have
various effect; they may both decrease and increase the probabilities
$\varphi_i$.

Now we consider the effects of selection and errors in more detail.
One well-known error is the one in background subtraction.  Every
cluster is located in a certain environment; thus the observed number
of galaxies in a given cluster has to be corrected for the expected
number of background galaxies.  Remember that Abell (1958) estimated a
local background for each cluster, while ACO used an universal
luminosity function.  This correction is subject to errors: in a
high-density environment the background may increase the observed
cluster richness $\cal R$ so that a poor cluster with ${\cal R < R}_0$
is observed as ${\cal R \geq R}_0$.  Here ${\cal R}_0$ is the minimal
richness of the cluster to be included in the catalog; for Abell
clusters of richness class 0 ${\cal R}_0=30$.  This systematic effect
is enhanced in superclusters where clusters are affected not only by
the universal background but also by overlapping clusters.  This
effect is detected in the Abell catalog; it makes the
two-dimensional correlation function (calculated for different radial
and angular separations) elongated in the radial direction and
increases the correlation length (Sutherland 1988).  To avoid such a
systematic error the background correction was made very carefully for
the APM cluster catalog (D97).  Since the background correction is
always influenced by random errors, the true minimal number of
galaxies in clusters ${\cal R}_{0t}$ fluctuates around the accepted
number ${\cal R}_0$.

Another well-studied selection effect is the dependence of the
observed number of clusters on Galactic latitude, $b$. Usually this
effect has been taken into account only by exclusion of clusters at
low Galactic latitudes, $b \leq 30^{\circ}$ (Sutherland 1988).  Our
analysis has shown that actually the spatial density of clusters is an
almost linear function of $\sin b$, even at $|b| > 30^{\circ}$.
This dependence can be described through the probability function:
\begin{equation}
\varphi_b = \cases{0, &$\sin b \le \sin b_0$;\cr\cr
{(\sin b - \sin b_0) \over (1-\sin b_0)}, &$\sin b > \sin b_0$;}
\label{eq7}
\end{equation}
here $\sin b_0$ is the value of $\sin b$ where the density of clusters
reaches zero.  Such a dependence was found by E97b, E97d, and in Paper
II, and is confirmed by the present study.  For Abell and APM cluster
samples investigated in this study the values of the parameter $\sin
b_0$ are given in Table~1.  The latitude dependence can be considered
as a random exclusion of faint and/or low surface brightness galaxies
from clusters, so that the number $\cal R$ decreases and a cluster may
fall below the limiting number $\cal R_0$, so that the cluster is not
included to the catalog.  From statistical point of view clusters are
included to the catalog only with a certain probability which
decreases with the increase of angular distance from Galactic
pole. This process decreases also the number of clusters in
superclusters; thus the density of clusters in rich superclusters
decreases with decreasing Galactic latitude more rapidly than the
density of all clusters.  This dependence can be modeled using mock
samples; random exclusion model with probability function (\ref{eq5})
describes well the latitude distribution of clusters in superclusters
of various richness (E97c).

The third selection effect is the dependence of the number of observed
clusters on the distance from the observer.  The mean density is
usually found by appropriate smoothing of the observed number of
clusters as a function of $z$ (Efstathiou \etal 1992, Croft \etal
1997).  The distance dependence of the spatial density of the cluster
sample is due to several different effects.  One effect is related to
the difficulty to detect, especially a poor cluster, on large
distances.  Secondly, nearby clusters cover too large an area on the sky
to use standard cluster searching algorithm, and are thus omitted.
Further, near the limit of the sample errors of the estimated redshift
influence the completeness of the sample. And finally, far from all
clusters have their redshifts measured.

The physical background of the difficulty to find a distant cluster is
not clear, but probably it is due to detection problems of galaxies of
low surface brightness at large distance.  This reduces the number of
galaxies found in the cluster below the threshold of ${\cal R}_0$.  A
similar detection problem may be the reason for the latitude
dependence of the spatial density of clusters.  Formally this
selection effect can be modeled as a random exclusion of clusters from
a general sample of true clusters.  Following E97b and E97d, we use
the linear probability function:
\begin{equation}
\varphi_{r1}(r) = d_0 - d_1 {r \over r_{lim}}, 
\label{eq8}
\end{equation}
where $d_0$ is the probability to detect a cluster in our vicinity
($r=0$); $d_0 - d_1$ is the probability to detect a cluster at
distance $r_{lim}$; and $r_{lim}$ is the limiting radius of the sample.
We adopt $d_0=1$ for the southern part of the Abell sample.  Table~1
lists the parameters $d_0$ and $d_1$ for northern and southern Abell
cluster samples.

The second distance-dependent selection effect is due to random errors
of estimated redshifts.  Clusters near the far limit of the sample are
selected by the following criterion: both the Abell and APM samples
have been selected using an upper limit for the estimated redshift,
$z_{lim}$.  Estimated redshifts are subject to random errors, with
approximately log-normal distribution of dispersion 1.3 in $\log z$
(see Figure~2).  This effect causes a deficit of clusters near the
limit $z \leq z_{lim}$.  On the other hand, some clusters within the
limit $z_{lim}$ have true redshifts exceeding the estimated one, thus
the density distribution has a large tail of clusters with $z_{meas} >
z_{lim}$ but $z_{est} < z_{lim}$.  To decrease this distance selection
effect we apply for the Abell sample an upper limit for the estimated
redshift, $z=0.15$, which exceeds by $\Delta z=0.02$ the redshift
limit of the sample used for our structure analysis, $z_{lim}=0.13$.
Figure~3 shows that a rapid decrease of the spatial density of
clusters near the limit of the sample can be avoided by this
procedure.  A similar cut would limit the APM sample to
$z_{lim}=0.10$.  Instead we applied an identical cut for both cluster
samples, $r_{lim} = 350$~\Mpc, at the cost of introducing this
selection effect in the APM sample.  The rapid decrease of the density
of APM clusters beyond 300~\Mpc\ is due to the sharp cutoff of the
estimated redshift at $z=0.118$ or $r=325$~\Mpc\ (see Figure~2).

The third distance-dependent selection effect is due to difficulty to
detect a very nearby cluster spanning a large area on the sky (D97).
This effect is practically absent in the Abell catalog; only the
Virgo cluster was excluded from the Abell catalog for this reason.
For APM clusters this effect is enhanced by the absence in the APM
sample galaxies brighter than $b_j = 17$ (which are overexposed and
cannot be measured automatically, see Maddox, Efstathiou \& Sutherland
1990).  Due to this selection effect the APM cluster sample is very
sparse at small distances of $r < 100$~\Mpc\ from us; this explains
also the absence of nearby rich superclusters in the APM sample, (see
Figure~3). 

So far we have discussed selection effects in samples of all clusters.
As not all clusters have a measured redshift, another
distance-dependent selection effect occurs in samples of clusters with
only measured redshifts.  The overall distance-dependent selection
probability was calculated by Efstathiou \etal (1992) and Croft \etal
(1997) by smoothing the observed redshift distribution.  We use a
different approach here, and calculate the combined effect of all
distance-dependent selection effects for Abell clusters by
eq.~(\ref{eq6}).  
For the APM cluster samples the combined effect of distance-dependent
selections can be approximated by the following law:
\begin{equation}
\varphi_{r2}(r) = \cases{0, &$r\le r_{min}$;\cr\cr
f_0{(r - r_{min}) \over (r_{max} - r_{min})}, &$r_{min} <r \leq r_{max}$;
\cr\cr
f_0{(r_{lim} - r) \over (r_{lim} - r_{max})}, &$r_{max} <r \leq r_{lim}$;
} 
\label{eq9}
\end{equation}
here we assume that at a distance $r_{max}$ clusters are detected with
a probability $f_0$.  This probability function was used to calculate
the comparison Poisson sample for the APM clusters. Values of the
parameters $r_{min}$, $r_{max}$, and $f_0$ are given in  Table~1.

We summarize the discussion of selection effects in Abell and APM
cluster catalogs as follows.  Both cluster catalogs are subject to
similar selection effects in Galactic latitude. The error due to
overlapping of clusters in superclusters is present only in the Abell
catalog.  This error may distort the structure of individual
superclusters (and the cluster correlation function on small
separations). However, it is of less importance for the study of the
structure of the supercluster-void network, since it does not decrease
systematically the number of clusters in superclusters.
Distance-dependent selection effects are rather small in the Abell
cluster sample, but very large in the APM sample.  We shall discuss
the influence of these selection effects on the study of the
supercluster-void network in section 5.5 below.

\subsection{Spatial density and volume covered by the Abell and APM
cluster samples}

In order to calculate the spatial density of Abell and APM clusters of
galaxies we generate Poisson samples with selection function
parameters as given in Table~1. We count the total number of
particles, and the number of particles left after the selection
functions have been applied.  We obtain $\rho = 27.6 \times
10^{-6}~h^3$ ~Mpc$^{-3}$ for the ACO.A1 sample and $\rho = 83 \times
10^{-6}~h^3$ ~Mpc$^{-3}$, for APM.A1.  Our value for Abell clusters
confirms earlier density estimate by E97d, but is about twice the
value found by Bahcall \& Cen (1993). The reason for this discrepancy
is our more detailed account of the selection function (see
eq.~(\ref{eq7})). 

The surface of the sky covered by the sample of Abell clusters, can be
estimated, when we fix the effective Galactic latitude limit of the
survey.  The large-scale distribution of high-density regions can be
investigated if superclusters of richness $N_{cl} \ge 4$ are included
in the sample (E97d). Figure~4 shows that the density of clusters in
rich superclusters drops to zero at $\sin b \approx 0.30$. Thus we
find that the Abell cluster survey covers 8.8 steradians (sr) on the
sky.  The APM survey is limited by $-0.30 \ge \sin\delta \ge -0.95$ in
declination and $\Delta RA = 130^{\circ}$ in right ascension,
corresponding to an area of 1.47~sr. The volume occupied by the Abell
sample ACO.A1 is $126 \times 10^6~h^{-3}$~Mpc$^3$. The depth of the
APM sample is from 50~\Mpc\ to 350~\Mpc, corresponding to a volume of
the APM survey $21 \times 10^6~h^{-3}$~Mpc$^3$.  Here we are speaking
on  the volume of the {\em sample} of Abell or APM clusters,  not the
volume occupied by Abell or APM clusters themselves.

We can also find the effective volume of both samples, defined by the
number of clusters in the sample divided by the mean density (we mean
the true mean density which is calculated after the correction for all
selection effects).  We get $60 \times 10^6~h^{-3}$~Mpc$^3$ and $11
\times 10^6~h^{-3}$~Mpc$^3$, for the ACO.A1 and APM.A1 samples. The
effective volumes of samples ACO.R1 and APM.R1 are even less, $40
\times 10^6~h^{-3}$~Mpc$^3$ and $5 \times 10^6~h^{-3}$~Mpc$^3$,
respectively.  We see that the selection effects reduce the effective
volume with respect to the actual volume of samples.  Volume estimates
show that the APM sample of all clusters occupies a volume about 1/6
of the volume of the corresponding Abell sample ACO.A1.  The volume
estimate given by T98 is larger than found in the present paper,
since T98 also included regions where the sample is very sparse.

\subsection{Correlation functions}

The cluster correlation function has been a subject of intensive
studies starting from the pioneering work by Klypin \& Kopylov (1983)
and Bahcall \& Soneira (1983).  In most cases the main purpose was the
determination of the correlation length, $r_0$, and the power index,
$\gamma$, of the function; for recent studies we refer to Sutherland
(1988), Efstathiou \etal (1992), Dalton \etal (1992), and Croft \etal
(1997).  These authors have shown that the correlation length for APM
clusters is smaller than that for Abell clusters.  This can be
explained in part by the overlapping of Abell clusters in
superclusters; a small intrinsic difference is also possible.  As
shown by real data and numerical simulations, the cluster correlation
function depends on cluster richness (Bahcall \& West 1992, Bahcall \&
Cen 1992, Frisch \etal 1995).  We also find a weak difference of the
cluster correlation length for Abell and APM samples. A much stronger
dependence exists for clusters located in different environments: the
correlation length for clusters in rich and very rich superclusters is
much higher than for all clusters (E97b, Paper II).  This effect is
due to variable density threshold in the definition of clusters in a
different environment, for an analysis of this effect see Einasto
\etal (1999b).  In the present study we confirm this result (see
Table~1): {\em there is no unique value of the cluster correlation length
-- it depends on cluster richness and environment}.

In the present study we have concentrated on the study of cluster
correlations at large separations.  For small separations the
correlation function characterizes the distribution of clusters in
superclusters, and for large separations the distribution of
superclusters themselves, i.e. properties of the supercluster-void
network (E97c).  The present study has confirmed earlier findings of
E97b, that on large scales the Abell cluster correlation function is
oscillating, i.e. it consists of alternating secondary maxima and
minima.  Oscillations are clearly detected if we use samples of
clusters in very rich superclusters. There is practically no
difference between the oscillatory behavior of samples of all clusters
and clusters with measured redshifts, ACO.A8 and ACO.R8, respectively.
We find a period of oscillations of $116 \pm 20$~\Mpc, very close to
the value found earlier.  The APM cluster sample also shows signs of
oscillations with the same period, if we use all clusters (sample
APM.A8).  If we use clusters with measured redshifts only (sample
APM.R8), we see again a strong secondary maximum, but at much larger
separation corresponding to the mutual distance of two dominating
superclusters in the sample, the Horologium-Reticulum and the Sculptor
superclusters (see Figure~6).  This is the main difference between
Abell and APM samples of clusters.  We continue the discussion of this
difference in the next Section.

\subsection{ Abell and APM clusters as tracers of the
supercluster-void network}

The principal statistic to characterize the distribution of matter on
large scales has been the power spectrum of clusters of galaxies.  Our
study confirms previous evidence (E97a, R98, and T98) for the
existence of a real differences between power spectra of Abell and APM
cluster samples.  The power spectrum of the Abell sample has a peak on
a scale of $\approx 120$~\Mpc.  A peaked power spectrum corresponds to
an oscillating correlation function (E97a) with a period of 120~\Mpc.
The APM cluster sample as analyzed by T98 has no feature on this
scale.  It has been argued that the APM cluster sample is free of the
projection and selection biases known to affect the Abell cluster
sample (T98).  For this reason the reality of the feature seen in the
Abell sample has been questioned (Postman 1998).  This brings us to
the central problem of our study: How well do the Abell and APM
cluster samples trace the structure of the Universe on large scales?

Our study has shown that selection effects are more complicated than
assumed previously.  In particular, the selection function in distance
depends on four different effects: the difficulty to detect clusters
at large and small distances, the deficiency of clusters at large
distance due to random errors in estimated redshifts used for the
selection of clusters, and the selection caused by observing programs
of cluster redshifts.  Our analysis has shown that the sample of all
Abell clusters (ACO.A1) is affected by distance-dependent selection
effects only on very large distances; this selection effect may be
reduced if the sample is cut at a redshift of $z=0.13$.  In contrast,
the APM sample with measured redshifts (APM.R1) is affected by several
distance-dependent selection effects within the range of distances of
interest for the present study.

T98 has compensated distance-dependent selection effects using weights
for clusters inversely proportional to the mean spatial density of the
cluster sample at its respective distance, following the prescription
by Feldman, Kaiser \& Peacock (1994, hereafter FKP).  T98 has also
applied truncation of the sample at large distances, $r_{lim}=600$~
and 400~\Mpc. The resulting estimates of the spectrum are similar, but
the scatter of the power spectrum is smaller when cutting the sample
at smaller distance.  T98 argue that the sample of APM clusters with
measured redshifts is representative for a volume of depth 400~\Mpc.

The Figure~3 shows that the spatial density of the APM sample is very
low at large distances.  The same effect can be seen also in Figure~1
of T98 (note that T98 do not plot the spatial density, but the number
of clusters in respective bins). The comparison of the spatial
densities of the Abell and APM samples in Figure~3 demonstrates that
the APM sample with measured redshifts becomes very sparse or diluted
at distances $r > 300$~\Mpc; in the Abell sample dilution becomes
strong only at distances $r > 400$~\Mpc. The question is: Is the use
of weights sufficient to compensate for the strong dilution observed
in the APM sample at large distance?

FKP argue that statistical properties of large scale structures can be
fixed by using sparse samples if density perturbations are Gaussian.
The Gaussian character of small-scale perturbations has been checked
by the analysis of FKP.  On large scales the problem is unsolved.
Szalay (1998) has demonstrated that the distribution of large
structures (supercluster-void network generated by Voronoi
tessellation) may be destroyed completely by randomization of the
phases of perturbations without changing the power spectrum.  A sparse
sample cannot distinguish between a regular supercluster-void network
and random distribution of high-density regions.  This simple example
shows very clearly that a very sparse sample is not suitable to
describe the real supercluster-void network.

Our experience has shown (Einasto \etal 1991, Frisch \etal 1995, E97c)
that dilution is not dangerous to locate high-density regions, as long
as the main structural elements are not destroyed.  The distribution
of matter on large scales is dominated by the supercluster-void
network.  Thus, in order to investigate the character of the mass
distribution on large scales, rich superclusters must be present in
sufficient quantities (E97c).  Figure~3 shows that the Abell cluster
sample is complete enough to trace the network of rich superclusters
up to $r_{lim} = 375$~\Mpc. The APM sample with measured redshifts
contains only two rich supercluster complexes within a distance of $r
= 325$~\Mpc.  The mutual separation of these two supercluster
complexes determines the shape of the correlation function and the
power spectrum of the whole APM sample of clusters with measured
redshifts. But this is not sufficient to trace the whole
supercluster-void network.  The sample is diluted and not all rich
superclusters actually present can be traced in this distance range
(see Figure~6).

This example shows that the use of the weights does not compensate for
the lack of data.  The APM sample of all clusters contains a
sufficiently large number of rich and very rich superclusters, but
most of the clusters in these superclusters do not yet have measured
redshifts.  The analysis of the sample of all APM clusters has shown
that using this larger sample will probably yield a more
representative picture of the supercluster-void network.  T98 has used
the APM cluster sample B for which a limit of estimated redshifts
$z_{est} \le 0.118$ was applied, corresponding to the apparent
magnitude limit of $m_X \le 19.4$.  This limit is implied by  the
limiting magnitude of the APM galaxy survey, $b_J= 20.5$, and the
magnitude range $[m_X - 0.5, m_X + 1.0]$ to determine the cluster
richness ${\cal R}$. Croft \etal (1997) have formed a deeper APM
cluster sample C, using a fainter limiting magnitude of the galaxy
catalog, $b_J = 21.0$, and a narrower magnitude range $m_X - 0.5,
m_X + 0.7]$, which yields $r_{lim} = 850$~\Mpc\ for the sample C.
This sample contains several rich superclusters at a distance of
$\approx 600$~\Mpc; these superclusters are also seen in the Abell
catalog as clusters of distance class 6 (Croft \etal 1997).
Unfortunately, this much deeper APM cluster sample has not yet been
studied to determine the power spectrum.

To conclude the discussion we can say that presently the Abell cluster
sample yields more accurate data on the structure of the
supercluster-void network than the APM sample with measured redshifts
-- the effective volume of the last sample is too small.

\section {Conclusions}

We have compared the spatial distribution of Abell and APM clusters
and cluster-defined superclusters in order to understand the
similarities and differences between the correlation functions and
power spectra of these clusters.  Our main results can be summarized
as follows.

1) We have compiled a catalog of superclusters on the basis of APM
clusters; the catalog contains data on 55 superclusters with at
least 4 members, it is given in Appendix; most clusters have only
estimated redshifts.

2) Abell and APM clusters of galaxies show almost identical
high-density regions (i.e.\ rich and very rich superclusters) in the
space where samples overlap, if all clusters are used to trace the
structure.

3) The sample of APM clusters with measured redshifts covers a much
smaller volume in space than that of the Abell clusters.  Statistical
properties of the APM sample with measured redshifts reflect the
distribution of clusters in this particular volume which is dominated
by two very rich supercluster complexes.  The Abell sample of clusters
contains 31 very rich superclusters and can be considered as a
candidate of a fair (representative) sample of the Universe.

4) The location of the secondary maximum of the correlation function
for APM clusters with measured redshifts at a separation of $r=185$\Mpc,
and the position of the maximum of the power spectrum, $k =
0.033$~\hmpc, correspond to the mutual separation between the
Horologium-Reticulum and the Sculptor superclusters; and do not
characterize the structure of the whole supercluster-void network. 

5) The analysis of the new Abell sample  of clusters confirms earlier
findings that the cluster power spectrum has a maximum on a scale of $k =
0.05$~\hmpc\  which corresponds to the period of the supercluster-void
network, 120~\Mpc.

6) The use of weights in calculation of the correlation function and
the power spectrum does not compensate for the lack of data.
Properties of the supercluster-void network can only be determined if
data are available for a sufficiently large number of rich
superclusters.

\section*{Appendix: APM supercluster catalog}

The catalog of superclusters of Abell clusters is based on cluster
sample ACO.A2, i.e. it contains all superclusters of richness class
$N_{cl} \geq 2$.  This catalog is published in Paper I.

Here we present a supercluster catalog based on the APM cluster
sample used in this paper.  The catalog is based on cluster sample
APM.A4, i.e. it contains all superclusters of richness class $N_{cl}
\geq 4$, the reason of the use of this limit was given above -- it is
due to the large number of clusters without measured redshifts in the
APM sample, thus increasing the limit of member clusters for the
catalog we hope to increase the reliability of the catalog.
$N_{cl}$ is the number of member clusters in the supercluster;
$RA_C$ and $\delta_C$ are coordinates of the center of the
supercluster (equinox 1950.0), derived from coordinates of individual
clusters; $D_C$ is the distance of the center from us; it follows the
list of Abell clusters which are members of the supercluster. An index
"e" after the Abell or APM cluster number in the column 6 shows that
this cluster has estimated velocity.  In the last column we list a
commonly used name of the supercluster, identifications show the
number of corresponding supercluster in the Table A1 of Paper I.

\acknowledgements We thank Enn Saar and Alexei Starobinsky for
stimulating discussion. This work was supported by the Estonian
Science Foundation grant 2625.  JE thanks Astrophysical Institute
Potsdam for hospitality where part of this study was performed.
HA thanks CONACyT for financial support under grant 27602-E.

{\tiny
\vskip-2cm
\begin{table*}
\tablenum{A1}
\begin{center}
\caption{The list of rich superclusters of APM clusters}
\begin{tabular}{rrrrrrrrrrrrrrrr}
\\
\hline
\\
(1)& (2) & (3) & (4) &(5) & &&&&(6)&&&&& & (7)  \\
\\
$No$ & $N_{CL}$ & $RA_C$ & $\delta_C$ & $D_C$ 
&&\multispan4 Abell-ACO No.&&&& &  \\
  & &  &  &\multispan1$ h^{-1} Mpc$ &&&&&& & &&&& \\

\\
\hline
\\
     1\rlap{$_c$} &     4 &     0.1 &   -62.0 &    292 &
   38\rlap{$_e$}  &
   47\rlap{$_e$}  &
  906\rlap{$_e$}  &
  932\rlap{$_e$}  &
  &
  &
  &
  &
  &
  &
  \cr
     2 &     7 &     1.1 &   -40.4 &    286 &
   2\rlap{$_e$} &
   6\rlap{$_e$} &
   9\rlap{$_e$} &
   25\rlap{$_e$} &
  32\rlap{$_e$} &
   944 &
   947 &
  &
  &
  &
9 (Sculptor region)
  \cr
     3 &    14 &     2.9 &   -34.0 &    304 &
    3 &
    10 &
    20 &
    35 &
    51 &
   57\rlap{$_e$} &
   58\rlap{$_e$} &
   59\rlap{$_e$} &
    72 &
    76 &
5 (Sculptor region)
 \cr
 & & & & &
   935 &
   937\rlap{$_e$} &
  942 &
  951\rlap{$_e$} &
   &
  &
  &
  &
  &
  &
  \cr
     4\rlap{$_c$} &     5 &     4.2 &   -51.3 &    315 &
   17\rlap{$_e$} &
   23\rlap{$_e$} &
  26\rlap{$_e$} &
    45 &
   81\rlap{$_e$} &
   &
  &
  &
  &
  &
  \cr
     5\rlap{$_c$} &     4 &     5.8 &   -51.8 &    250 &
   31\rlap{$_e$} &
 60\rlap{$_e$} &
 70\rlap{$_e$} &
     79 &
  &
  &
  &
  &
  &
  &
  \cr
     6 &    24 &    10.0 &   -26.7 &    306 &
 48\rlap{$_e$} &
 55\rlap{$_e$} &
 65\rlap{$_e$} &
 67\rlap{$_e$} &
   73 &
   80\rlap{$_e$} &
 82\rlap{$_e$} &
   87\rlap{$_e$} &
    92 &
    96 &
9 (Sculptor)
  \cr
 & & & & &
   99 &
   100 &
   101 &
   103\rlap{$_e$} &
  104 &
   110 &
   112 &
   119 &
  121\rlap{$_e$} &
   124 &
   \cr
 & & & & &
  127\rlap{$_e$} &
 128\rlap{$_e$} &
   130 &
 137\rlap{$_e$} &
  &
  &
  &
  &
  &
  &
  \cr
     7\rlap{$_c$} &     5 &    12.6 &   -50.2 &    319 &
 102\rlap{$_e$} &
 111\rlap{$_e$} &
 113\rlap{$_e$} &
 129\rlap{$_e$} &
 135\rlap{$_e$} &
  &
  &
  &
  &
  &
  \cr
     8 &    16 &    19.3 &   -35.5 &    312 &
   123 &
 131\rlap{$_e$} &
 134\rlap{$_e$} &
  144 &
 151\rlap{$_e$} &
 153\rlap{$_e$} &
 164\rlap{$_e$} &
 167\rlap{$_e$} &
 169\rlap{$_e$} &
 174\rlap{$_e$} &
22
   \cr
 & & & & &
 175\rlap{$_e$} &
 178\rlap{$_e$} &
 180\rlap{$_e$} &
 186\rlap{$_e$} &
  188\rlap{$_e$} &
  199\rlap{$_e$} &
 &
  &
  &
  &
  \cr
     9 &     4 &    19.5 &   -37.0 &    221 &
  158\rlap{$_e$} &
   160 &
   162 &
   173 &
  &
  &
  &
  &
  &
  &
23
  \cr
    10\rlap{$_c$} &     4 &    19.7 &   -24.8 &    304 &
  156\rlap{$_e$} &
  159\rlap{$_e$} &
 168\rlap{$_e$} &
  170\rlap{$_e$} &
 &
  &
  &
  &
  &
  &
  \cr
    11 &     4 &    23.5 &   -32.8 &    192 &
   182 &
   193 &
   194 &
   209 &
  &
  &
  &
  &
  &
  &
  \cr
    12 &     4 &    25.6 &   -55.0 &    257 &
   204 &
   211 &
   213 &
   214 &
  &
  &
  &
  &
  &
  &
  \cr
    13\rlap{$_c$} &     8 &    28.0 &   -33.8 &    313 &
  205\rlap{$_e$} &
  206\rlap{$_e$} &
 212\rlap{$_e$} &
  222\rlap{$_e$} &
 230\rlap{$_e$} &
  232\rlap{$_e$} &
 233\rlap{$_e$} &
  235\rlap{$_e$} &
  &
  &
  \cr
    14 &     7 &    30.4 &   -41.5 &    323 &
 215\rlap{$_e$} &
  229\rlap{$_e$} &
    234 &
   236 &
  237\rlap{$_e$} &
  240\rlap{$_e$} &
  242\rlap{$_e$} &
 &
  &
  &
37
  \cr
    15\rlap{$_c$} &     4 &    32.5 &   -40.4 &    285 &
  228\rlap{$_e$} &
   246 &
 251\rlap{$_e$} &
  259\rlap{$_e$} &
   &
  &
  &
  &
  &
  &
  \cr
    16 &     9 &    34.2 &   -47.8 &    310 &
 243\rlap{$_e$} &
   249 &
  250\rlap{$_e$} &
  256\rlap{$_e$} &
  258\rlap{$_e$} &
  262\rlap{$_e$} &
    263 &
  264\rlap{$_e$} &
  267\rlap{$_e$} &
  &
37
  \cr
    17\rlap{$_c$} &     4 &    39.1 &   -33.5 &    307 &
  276\rlap{$_e$} &
  277\rlap{$_e$} &
  280\rlap{$_e$} &
  295\rlap{$_e$} &
  &
  &
  &
  &
  &
  &
  \cr
    18 &     4 &    41.3 &   -46.1 &    275 &
   289 &
   290 &
  297\rlap{$_e$} &
  306\rlap{$_e$} &
  &
  &
  &
  &
  &
  &
41
  \cr
    19 &     7 &    41.8 &   -24.9 &    309 &
  303\rlap{$_e$} &
  300\rlap{$_e$} &
  291\rlap{$_e$} &
  281\rlap{$_e$} &
   305 &
   309 &
   320 &
  &
  &
  &
43
  \cr
    20\rlap{$_c$} &     6 &    43.1 &   -41.1 &    318 &
  332\rlap{$_e$} &
  325\rlap{$_e$} &
  321\rlap{$_e$} &
  316\rlap{$_e$} &
  299\rlap{$_e$} &
   293 &
  &
  &
  &
  &
  \cr
    21\rlap{$_c$} &     5 &    43.6 &   -51.3 &    323 &
  301\rlap{$_e$} &
  324\rlap{$_e$} &
  326\rlap{$_e$} &
  327 &
   328 &
  &
  &
  &
  &s
  &
  \cr
    22 &    30 &    46.9 &   -53.7 &    201 &
 285 &
  287\rlap{$_e$} &
 296\rlap{$_e$} &
  298\rlap{$_e$} &
 302\rlap{$_e$} &
   304 &
  308\rlap{$_e$} &
  310\rlap{$_e$} &
  314\rlap{$_e$} &
  323 &
48 (Horologium-Reticulum)
  \cr
 & & & & &
  330\rlap{$_e$} &
  340\rlap{$_e$} &
  343\rlap{$_e$} &
 346\rlap{$_e$} &
  356 &
   364 &
   380 &
  382\rlap{$_e$} &
  383\rlap{$_e$} &
  389\rlap{$_e$} &
  \cr
 & & & & &
  391 &
   395 &
   396 &
   397 &
   399 &
   403 &
   421 &
   434 &
   445 &
  463\rlap{$_e$} &
  \cr
 & & & & &
  \cr
    23\rlap{$_c$} &     7 &    49.5 &   -44.5 &    318 &
  350\rlap{$_e$} &
   351 &
  384\rlap{$_e$} &
 385\rlap{$_e$} &
  390\rlap{$_e$} &
 392\rlap{$_e$} &
  393\rlap{$_e$} &
  &
  &
  &
  \cr
    24 &     9 &    50.4 &   -43.0 &    179 &
   342 &
  358\rlap{$_e$} &
   360 &
   365 &
   387 &
   388 &
   413 &
   415 &
   443 &
  &
48
  \cr
    25 &    16 &    51.1 &   -44.3 &    209 &
   357 &
   362 &
   366 &
   367 &
   369 &
   370 &
 372\rlap{$_e$} &
   373 &
   374 &
   377 &
48
  \cr
 & & & & &
   400 &
  408\rlap{$_e$} &
   433 &
   450 &
  464\rlap{$_e$} &
  478 &
  &
  &
  &
  &
  \cr
    26 &    23 &    52.3 &   -32.5 &    294 &
 375\rlap{$_e$} &
  376\rlap{$_e$} &
  378\rlap{$_e$} &
  379\rlap{$_e$} &
  381\rlap{$_e$} &
  394\rlap{$_e$} &
  401\rlap{$_e$} &
 402\rlap{$_e$} &
  406\rlap{$_e$} &
   407 &
53 (Fornax-Eridanus)
 \cr
 & & & & &
  409\rlap{$_e$} &
  412\rlap{$_e$} &
  414 &
   416 &
   418 &
  419\rlap{$_e$} &
  422\rlap{$_e$} &
  423 &
   424 &
  426\rlap{$_e$} &
   \cr
 & & & & &
 428 &
   429 &
  441\rlap{$_e$} &
 &
  &
  &
  &
  &
  &
  &
  \cr
    27\rlap{$_c$} &    15 &    61.1 &   -29.9 &    305 &
 457\rlap{$_e$} &
  458\rlap{$_e$} &
  460\rlap{$_e$} &
  461\rlap{$_e$} &
  466\rlap{$_e$} &
  467 &
  471\rlap{$_e$} &
 472\rlap{$_e$} &
  482\rlap{$_e$} &
  487\rlap{$_e$} &
   \cr
 & & & & &
  488\rlap{$_e$} &
 491\rlap{$_e$} &
  493 &
  499\rlap{$_e$} &
  503\rlap{$_e$} &
  &
  &
  &
  &
  &
  \cr
    28\rlap{$_c$} &     5 &    69.4 &   -26.0 &    278 &
    498 &
 507\rlap{$_e$} &
  526\rlap{$_e$} &
  535\rlap{$_e$} &
  537\rlap{$_e$} &
  &
  &
  &
  &
  &
  \cr
    29\rlap{$_c$} &    11 &    69.6 &   -46.2 &    317 &
  501\rlap{$_e$} &
  512\rlap{$_e$} &
  514\rlap{$_e$} &
  521\rlap{$_e$} &
  522\rlap{$_e$} &
  523\rlap{$_e$} &
 528\rlap{$_e$} &
  529\rlap{$_e$} &
  534\rlap{$_e$} &
543\rlap{$_e$} &
  \cr
 & & & & &
544\rlap{$_e$} &
   &
  &
  &
  &
  &
  &
  &
  &
  &
  \cr
    30\rlap{$_c$} &     4 &    74.3 &   -29.6 &    283 &
 539\rlap{$_e$} &
  561\rlap{$_e$} &
  563\rlap{$_e$} &
  564\rlap{$_e$} &
  &
  &
  &
  &
  &
  &
  \cr
    31\rlap{$_c$} &     7 &    76.7 &   -43.2 &    305 &
  546\rlap{$_e$} &
  565\rlap{$_e$} &
  570\rlap{$_e$} &
  576\rlap{$_e$} &
 580\rlap{$_e$} &
   594\rlap{$_e$} &
  598\rlap{$_e$} &
 &
  &
  &
  \cr
    32\rlap{$_c$} &     4 &    77.5 &   -29.2 &    316 &
  568\rlap{$_e$} &
  569\rlap{$_e$} &
  581\rlap{$_e$} &
  599\rlap{$_e$} &
  &
  &
  &
  &
  &
  &
  \cr
    33 &     7 &    77.9 &   -41.4 &    225 &
  575\rlap{$_e$} &
  577\rlap{$_e$} &
  578 &
   583 &
   585 &
   593 &
   601 &
  &
  &
  &
65
  \cr
    34\rlap{$_c$} &     4 &    78.2 &   -42.4 &    272 &
  584\rlap{$_e$} &
  588\rlap{$_e$} &
  596\rlap{$_e$} &
  597\rlap{$_e$} &
  &
  &
  &
  &
  &
  &
  \cr
    35\rlap{$_c$} &     4 &   311.2 &   -32.4 &    313 &
  606\rlap{$_e$} &
   608 &
  610\rlap{$_e$} &
  612\rlap{$_e$} &
 &
  &
  &
  &
  &
  &
  \cr
    36\rlap{$_c$} &     4 &   315.8 &   -35.5 &    324 &
    618 &
  622\rlap{$_e$} &
  628\rlap{$_e$} &
 632\rlap{$_e$} &
  &
  &
  &
  &
  &
  &
  \cr
    37\rlap{$_c$} &    11 &   318.0 &   -42.7 &    313 &
  619\rlap{$_e$} &
  624\rlap{$_e$} &
  627\rlap{$_e$} &
   630 &
  633\rlap{$_e$} &
  639\rlap{$_e$} &
  640\rlap{$_e$} &
  649\rlap{$_e$} &
  652\rlap{$_e$} &
 655\rlap{$_e$} &
  \cr
 & & & & &
  667\rlap{$_e$} &
 &
  &
  &
  &
  &
  &
  &
  &
  &
  \cr
    38 &     8 &   318.8 &   -45.8 &    275 &
  634\rlap{$_e$} &
  635\rlap{$_e$} &
  641\rlap{$_e$} &
   642 &
   650 &
   653 &
   657 &
   659 &
  &
  &
183
  \cr
    39\rlap{$_c$} &     7 &   319.1 &   -43.9 &    214 &
  625\rlap{$_e$} &
 637\rlap{$_e$} &
    644 &
 645\rlap{$_e$} &
  647\rlap{$_e$} &
   658\rlap{$_e$} &
  681\rlap{$_e$} &
 &
  &
  &
  \cr
    40\rlap{$_c$} &     4 &   325.4 &   -51.5 &    315 &
   689 &
  691\rlap{$_e$} &
  698\rlap{$_e$} &
  705\rlap{$_e$} &
  &
  &
  &
  &
  &
  &
  \cr
    41 &     4 &   325.8 &   -42.2 &    180 &
   688 &
   700 &
   709 &
   711 &
  &
  &
  &
  &
  &
  &
182
  \cr
    42 &     5 &   326.8 &   -32.1 &    257 &
   674 &
  693\rlap{$_e$} &
  714\rlap{$_e$} &
   721 &
   749 &
  &
  &
  &
  &
  &
190
  \cr
    43 &     8 &   328.6 &   -19.1 &    305 &
  704\rlap{$_e$} &
  718\rlap{$_e$} &
   723\rlap{$_e$} &
 728\rlap{$_e$} &
   735 &
 737\rlap{$_e$} &
   741 &
  745\rlap{$_e$} &
  &
  &
  \cr
    44\rlap{$_c$} &     4 &   328.7 &   -57.3 &    308 &
  712\rlap{$_e$} &
 719\rlap{$_e$} &
  730\rlap{$_e$} &
  736\rlap{$_e$} &
  &
  &
  &
  &
  &
  &
  \cr
    45 &     4 &   331.4 &   -69.6 &    192 &
   731 &
  739\rlap{$_e$} &
   740 &
  756\rlap{$_e$} &
  &
  &
  &
  &
  &
  &
  \cr
    46 &    14 &   332.8 &   -56.7 &    216 &
  685\rlap{$_e$} &
  695\rlap{$_e$} &
   708 &
  717\rlap{$_e$} &
   726 &
   732 &
  744\rlap{$_e$} &
   766 &
   776 &
   777 &
192 
  \cr
 & & & & &
   780 &
  788\rlap{$_e$} &
  790\rlap{$_e$} &
   792 &
  &
  &
  &
  &
  &
  &
  \cr
    47 &    10 &   336.4 &   -48.8 &    295 &
   754 &
  769\rlap{$_e$} &
  770\rlap{$_e$} &
  773\rlap{$_e$} &
   778 &
  781\rlap{$_e$} &
  782 &
   787 &
   793\rlap{$_e$} &
 796\rlap{$_e$} &
197 (Grus)
  \cr
 & & & & &
  \cr
    48\rlap{$_c$} &     6 &   338.7 &   -37.9 &    312 &
 794\rlap{$_e$} &
  795\rlap{$_e$} &
  797\rlap{$_e$} &
  798\rlap{$_e$} &
  799\rlap{$_e$} &
  803\rlap{$_e$} &
  &
  &
  &
  &
  \cr
    49 &     4 &   341.1 &   -45.2 &    255 &
   811 &
   812 &
   814 &
   815 &
  &
  &
  &
  &
  &
  &
197 (Grus)
  \cr
    50 &     5 &   342.3 &   -64.6 &    272 &
    774 &
   813 &
   822 &
 837\rlap{$_e$} &
  849\rlap{$_e$} &
  &
  &
  &
  &
  &
200
  \cr
    51 &    10 &   345.2 &   -30.8 &    304 &
   827 &
  831\rlap{$_e$} &
  832\rlap{$_e$} &
  836\rlap{$_e$} &
  840 &
   847\rlap{$_e$} &
 853\rlap{$_e$} &
   864 &
  868\rlap{$_e$} &
  869\rlap{$_e$} &
209
  \cr
 & & & & &
  \cr
    52 &     4 &   348.9 &   -42.1 &    309 &
  861\rlap{$_e$} &
   878 &
   883 &
  885\rlap{$_e$} &
  &
  &
  &
  &
  &
  &
  \cr
    53 &    12 &   349.8 &   -22.9 &    311 &
    862 &
   866 &
  876\rlap{$_e$} &
 880\rlap{$_e$} &
   881\rlap{$_e$} &
 887\rlap{$_e$} &
  888\rlap{$_e$} &
   889\rlap{$_e$} &
  890 &
  894\rlap{$_e$} &
209
   \cr
 & & & & &
  896\rlap{$_e$} &
  903\rlap{$_e$} &
 &
  &
  &
  &
  &
  &
  &
  &
  \cr
    54\rlap{$_c$} &     4 &   352.6 &   -32.5 &    294 &
  897\rlap{$_e$} &
  899\rlap{$_e$} &
  912\rlap{$_e$} &
  913\rlap{$_e$} &
  &
  &
  &
  &
  &
  &
  \cr
    55 &     7 &   358.9 &   -34.2 &    147 &
     1 &
     5 &
    12 &
   905 &
   933 &
   938 &
   945 &
  &
  &
  &
220 (10 - Pisces-Cetus)
  \cr

\hline
\label{tab:sc}
\end{tabular}
\end{center}
\end{table*}
}


\begin{thebibliography}{}

\bibitem{a1} Abell, G., 1958, \apjs, 3, 211

\bibitem{a2}  Abell, G., Corwin, H., Olowin, R., 1989, \apjs,
70, 1 (ACO)

\bibitem{and2} Andernach, H., \& Tago, E., 1998, Proc.  "Large Scale
Structure: Tracks and Traces", eds.  V.~M\"uller, S.\, Gottl\"ober,
J.P.\, M\"ucket, \& J.\, Wambsganss, World Scientific, Singapore, p.~147

\bibitem{bah} Bahcall, N.A., 1988, ARA\&A, 26, 631  

\bibitem{bc92} Bahcall, N.A., \& Cen, R., 1992, ApJ, 398, L81

\bibitem{b2} Bahcall N.A. \& Cen R., 1993, \apj 407, L49 

\bibitem{bs83} Bahcall, N.A. \& Soneira, R. 1983, ApJ, 270, 20

\bibitem{bw92} Bahcall, N.A. \& West, M., 1992, ApJ, 392, 419

\bibitem{beks} Broadhurst, T.J., Ellis, R.S., Koo, D.C., \& Szalay, A.S.
  1990, Nature, 343, 726

\bibitem{cro} Croft, R.A.C., Dalton, G.B., Efstathiou, G.,
Sutherland, W.J., \& Maddox, S.J., 1997, \mn, 291, 305  

\bibitem{dm92} Dalton, G.B., Efstathiou, G., Maddox, S.J., \&
Sutherland, W.J., 1992, ApJ, 390, L1

\bibitem{dm97} Dalton, G.B., Maddox, S.J., Sutherland, W.J., \& Efstathiou,
  G. 1997, MNRAS, 289, 263

\bibitem{edsm92} Efstathiou, G., Dalton, G.B., Sutherland, W.J., \&
Maddox, S.J., MNRAS, 257, 125

\bibitem{e97b} Einasto, J., Einasto, M., Frisch, P., Gottl\"ober, S.,
M\"uller, V., Saar, V., Starobinsky, A.A., Tago, E., Tucker, D.,
Andernach, H., 1997b, \mn, 289, 801, (E97b)


\bibitem{e97c} Einasto, J., Einasto, M., Frisch, P., Gottl\"ober, S.,
M\"uller, V., Saar, V., Starobinsky, A.A., Tucker, D., 1997c, \mn,
289, 813, (E97c)

\bibitem{e97a} Einasto, J., Einasto, M., Gottl\"ober, S., M\"uller,
V., Saar, V., Starobinsky, A.  A., Tago, E., Tucker, D., Andernach,
H., Frisch, P., 1997a, Nature 385, 139 (E97a)

\bibitem{e91} Einasto, J.,  Einasto, M., Gramann, M., \& Saar, E.
1991, MNRAS, 248,  593 

\bibitem{e99a} Einasto, J., Einasto, M., Tago, E., Starobinsky, A.A.,
Atrio-Barandela, F., M\"uller, V., Knebe, A., Frisch, P., Cen, R.,
Andernach, H., \& Tucker, D., 1999a, \apj, {519}, 441

\bibitem{e99b} Einasto, J., Einasto, M., Tago, E., M\"uller, V.,  
Knebe, A., Cen, R., Starobinsky, A.A. \& Atrio-Barandela, F., 1999b,
\apj, {519}, 456

\bibitem{e80} Einasto J., J\~oeveer M.  \& Saar E., 1980, MNRAS,
193, 503 

\bibitem{me94}  Einasto, M., Einasto, J., Tago, E., Dalton, G.B.,
Andernach, H., 1994, MNRAS, 269, 301 (EETDA)

\bibitem{ein00} Einasto, M., Einasto, J., Tago, E., Hasinger, G.,
M\"uller, V., \& Andernach, H. 2001, ApJ, submitted (Paper I)   

\bibitem{me97}  Einasto, M., Tago, E., Jaaniste, J., Einasto, J., 
Andernach, H., 1997d, AAS, 123, 119 (E97d)

\bibitem{fkp} Feldman, H.A., Kaiser, N., \& Peacock, J.A., 1994, ApJ,
426, 23

\bibitem {f95} Frisch, P., Einasto, J., Einasto, M., Freudling, W., 
Fricke, K.J., Gramann, M., Saar, V., \& Toomet, O., 1995, AA, 296, 611

\bibitem{kk83} Klypin, A.A. \& Kopylov, A.A. 1983, Soviet Astr. Letters,
  9, 41

\bibitem{m90} Maddox, S.J., Efstathiou, G., \& Sutherland, W.J. 1990, MNRAS,
 246, 433

\bibitem{m97} Maddox, S.J., Efstathiou, G., \& Sutherland, W.J. 1996, MNRAS,
  283, 1227

\bibitem{m58} Mattig, W., 1958, Astr. Nachr. 284, 109

\bibitem{mil00} Miller, C., \& Batuski, D., 2000, ApJ, submitted
[astro-ph/0002295]
   
\bibitem{oort} Oort, J., 1983, ARA\&A, 21, 373

\bibitem{par} Park, C., \& Lee, S., 1998, subm.\ to J.\ Korean
Astron.\ Soc., [astro-ph/9809372]

\bibitem{pw92} Peacock J., \& West M.J., 1992, MNRAS 259, 494

\bibitem{press} Press, W. H., Teukolsky, S. A., Vetterling, W. T. \&
Flannery, B. P., {\em Numerical recipes in FORTRAN. The art of scientific
computing}, Cambridge University Press, 1992, 2nd ed.


\bibitem{pos} Postman, M., 1998, {\em Evolution of Large Scale
Structure: From Recombination to Garching}, eds.\ A.J.~Banday,
R.K.~Sheth, L.N.~da\, Costa, Garching 1999, p.\ 270, [astro-ph/9810088]

\bibitem{R98} Retzlaff, J., Borgani, S., Gottl\"ober, S., Klypin, A.,
  \& M\"uller, V. 1998, NewA 3, 631  (R98)

\bibitem{sut88} Sutherland, W.J., 1988, MNRAS, 234, 159 

\bibitem{sut99} Sutherland, W.J., Tadros, H., Efstathiou, G., Frenk,
C.S., Keeble, O., \& Maddox, S., McMahon, R.G., Oliver, S., Rowan-Robinson, M.,
    Saunders, W., \& White, S.D.M., 1999, MNRAS 308, 289 [astro-ph/9901189]


\bibitem{szalay} Szalay, A. S., 1998, in {\em Proceedings of 18th
Texas Symposium on Relativistic Astrophysics}, eds. A. Olinto, J.
Frieman, and D.  Schramm, World Scientific, Singapore, p. 136
 
\bibitem{t98} Tadros, H., Efstathiou, G. \& Dalton, G.  1998, MNRAS, 296, 995
   (T98)

\bibitem{t00} Tago, E., Einasto, J., Einasto, M., Hasinger, G.,
M\"uller, V., \& Andernach, H. 2001, (submitted to ApJ, Paper II)
  
\bibitem{v98} Vogeley, M. 1998, {The Evolving Universe},
ed. D. Hamilton, (Dordrecht: Kluwer), p. 395 [astro-ph/9805160]

\end{thebibliography}
\end{document}